%% file: paper.tex
\documentclass[final,5p,times,twocolumn]{elsarticle}

\usepackage[T1]{fontenc}
\usepackage[latin9]{inputenc}
\usepackage{amsmath}
\usepackage{amssymb}
\usepackage{graphicx}
\usepackage{xcolor}
\usepackage{hyperref}
\usepackage{listings}
\usepackage{booktabs}
\usepackage{subfig}

\def\binary{\texttt{binary64}}
\def\twosum{\texttt{TwoSum}}
\def\twoprod{\texttt{TwoProd}}
\def\blas{BLAS}

\def\exsum{{\sc exsum}}

\def\exdotp{{\sc exdot}}

\def\exgemv{{\sc exgemv}}
\def\extrsv{{\sc extrsv}}
\def\exgemm{{\sc exgemm}}

\def\CC{{ C\nolinebreak[4]\hspace{-.05em}\raisebox{.4ex}{\tiny\bf ++ }}}
\def\feltor{{\textsc{Feltor }}}

\begin{document}
\title{Reproducibility, accuracy and performance of the \feltor code and library on parallel computer architectures}

\author[cop]{Matthias Wiesenberger\corref{cor1}} \ead{mattwi@fysik.dtu.dk}
\author[tub,uibk]{Lukas Einkemmer} %\ead{einkemmer@na.uni-tuebingen.de}
\author[phys-uibk]{Markus Held}
\author[bsc]{Albert Gutierrez-Milla}
\author[bsc]{Xavier S\'aez}
\author[kth]{Roman Iakymchuk}

\address[cop]{Department of Physics, Technical University of Denmark (DTU), Denmark}
\address[tub]{Department of Mathematics, University of T\"ubingen, Germany}
\address[uibk]{Department of Mathematics, Universit\"at Innsbruck, Austria}
\address[phys-uibk]{Institute for Ion Physics and Applied Physics, Universit\"at Innsbruck, Austria}
\address[bsc]{Barcelona Supercomputing Center, Barcelona, Spain}
\address[kth]{Department of Computational Science and Technology, Royal Institue of Technology (KTH), Sweden}

\cortext[cor1]{Corresponding author}

\begin{abstract}
    \feltor is a modular and free scientific software package.
    It allows developing platform independent code that runs on a
    variety of parallel computer architectures ranging from laptop CPUs to
    multi-GPU distributed memory systems.
    \feltor consists of both a numerical library and a collection of application
    codes built on top of the library.
    Its main target are two- and three-dimensional drift- and gyro-fluid simulations
    with discontinuous Galerkin methods as the main numerical discretization technique.

    %First, we investigate accuracy and reproducibility.
    We observe that numerical simulations of a recently developed
    gyro-fluid model produce non-deterministic results in parallel computations.
    First, we show how we restore accuracy and bitwise reproducibility
    algorithmically and programmatically.
    In particular, we adopt an implementation of the exactly rounded dot product
    based on long accumulators, which avoids accuracy losses especially
    in parallel applications.
    However, reproducibility and accuracy alone fail to indicate correct
    simulation behaviour.
    In fact, in the physical model slightly different initial
    conditions lead to vastly different end states.
    This behaviour translates to its numerical representation.
    %argue that in ill-conditioned physical problems numerical
    %perturbations always grow exponentially.
    Pointwise convergence, even in principle, becomes impossible for long simulation
    times.
    We briefly discuss alternative methods to ensure the correctness of results
    like the convergence of reduced physical quantities of interest,
    ensemble simulations, invariants or reduced simulation times.

    In a second part, we explore important performance tuning considerations.
    %in our implementation of the aforementioned algorithms and equations.
    We identify latency and memory bandwidth as the main performance indicators
    of our routines.
    Based on these, we propose a parallel performance model
    that predicts the execution time of algorithms implemented in \feltor and
    test our model on a selection of parallel hardware architectures.
    We are able to predict the execution time % of algorithms
    with a relative error of less than $25\%$
    for problem sizes between $10^{-1}$ and $10^3$ MB.
    Finally, we find that the product of latency and bandwidth gives a minimum
    array size per compute node to achieve a scaling efficiency above $50\%$
    (both strong and weak).

\end{abstract}

\maketitle

\input{introduction.tex}
\input{feltor.tex}

\input{reproducibility.tex}
%\input{numerical-method.tex}
%\input{comparison.tex}
\input{performance.tex}
\input{conclusion.tex}

\bibliography{paper}{}
\bibliographystyle{plain}

\end{document}

%% file: introduction.tex
\section{Introduction}
For the description of low-frequency dynamics in magnetized plasmas
drift-reduced Braginskii (also called \textit{drift-fluid})~\cite{Braginskii,
Zeiler1997,Madsen2016} and gyro-fluid
models~\cite{Beer1996,Scott2010,Madsen2013,Held2016a} have been established.
Compared to kinetic descriptions the reduced dimensionality in these fluid
models significantly lowers the computational cost.  Furthermore, both of these
approaches remove the small time and spatial scales that are associated with
the gyration of charged particles in the magnetic field.
%~\cite{Braginskii, Zeiler1997, Scott2010, Garcia2006,Yan2013,Madsen2013,Halpern2016, Held2016a, Rasmussen2016, Ongena2016}.
%
Still, simulations of phenomena in magnetized plasmas are in general highly
challenging and require the use of advanced numerical algorithms as well as the
increasing power of high-performance computers~\cite{Fasoli2016}.

There are several codes implementing drift- and gyro-fluid models (among others
References~\cite{Tamain2010,Kendl2014,Halpern2016,Rasmussen2016}) and
Reference~\cite{Dudson2015} provides an actual framework for the implementation
of more general fluid equations.  In recent years code projects have focused on
the capability to efficiently invert nonlinear elliptic equations ( see for example
References~\cite{wiesenberger2014,Dudson2015,Halpern2016}).  This feature is
especially needed in models that avoid the so-called
\textit{Oberbeck-Boussinesq approximation} and do not distinguish between
fluctuating and background quantities.  For example in a gyro-fluid model we
have the nonlinear elliptic equation $\nabla\cdot(N\nabla_\perp\phi) = n-N$,
where $n$ is the electron density, $N$ is the ion gyro-center density, $\phi$
the electric potential and $\nabla_\perp$ the gradient in the direction
perpendicular to the magnetic field~\cite{Madsen2013}.  Current interest also
includes the implementation of the flux-coordinate independent approach to
discretize derivatives along arbitrary magnetic field
lines~\cite{Hariri2013,Stegmeir2014,Held2016}.  This type of scheme is
particularly important if a magnetic field aligned coordinate system is
unavailable due to singularities in the coordinate transformations (X-points).
It is then challenging to resolve the inherent anisotropy of the plasma
dynamics parallel and perpendicular to magnetic field lines.
Let us point out here that in the codes mentioned so far finite difference numerical methods are prevailing over more advanced schemes
and the efficient use of GPUs or other accelerator cards is largely absent.
However, accelerators have the potential to significantly improve performance
and codes that support them will be required to fully exploit the next
generation of supercomputers \cite{ascac}. Additionally, both GPUs and Intel
Xeon Phi co-processors are more energy efficient than conventional CPUs. Note
that energy consumption is the main challenge in building the next generation
of supercomputers.
%Thus, one part of this contribution is a thorough study of the performance characteristics
%of our own code \feltor on the aforementioned architectures.

Finally, we criticize that Reference~\cite{Dudson2015} is the only code
that can be classified as free software in our community, which
severely limits the possibility
to verify, interoperate with, reuse or even reproduce published results~\cite{Wilkinson2016}.
Reproducibility is of particular importance for parallel scientific
computations due to the non-deterministic nature of parallel computations that
increases even further with novel task-based approaches and dynamic scheduling
techniques. To address this issue or rather to ensure reproducibility, the top
publishers, journals, and conferences take very active initiatives. For
instance, the Supercomputing (SC) conference, the top conference in the field
of high-performance computing, makes it mandatory to provide an appendix
regarding reproducibility to be considered for "Best Paper" or "Best Student Paper"
under the SC Reproducibility Initiative~\cite{sc-repro-initiative}. The ACM
Transactions of Mathematical Software (TOMS) encourages authors to follow the
Replicated Computational Results (RCR) initiative~\cite{acm-toms-rcr}, meaning
the software is also reviewed in terms of replicating the presented results.
Furthermore, ACM introduced "Artifact Review and
Badging"~\cite{acm-repro-initiative}, which includes: \textit{repeatability}, when
the same team follows the same measurement procedure on the same experimental
setup; \textit{replicability}, when a different team measures the results on
the same experimental setup; \textit{reproducibility}, when a different team
measures the results on a different experimental setup. To that end, the issue
of non-reproducibility in a parallel environment is gathering attention and the
community aims to address it through various initiatives. Here, we join this
effort and tackle the non-reproducibility issue by a) making our software
publicly available, b) applying reproducible algorithmic solutions and c)
ensuring reproducibility from the programming perspective with the emphasis on
pitfalls and strengths of the environmental setup like compilers.

%Also, a framework that allows the usage of GPUs or other accelerator cards
%is absent from the literature.
To conclude the introduction let us here briefly mention the capabilities and background of our code.
\textsc{Feltor} is
a modular and free software package that we have developed
particularly for the use in full-F (no Oberbeck-Boussinesq approximation) drift- and gyro-fluid models~\cite{Held2016a,wiesenberger2017b,Kendl2017,Held2018}.
We use discontinuous Galerkin methods~\cite{Cockburn1998, Arnold2001} to spatially discretize model equations.
Our efforts to enable three-dimensional
simulations include the flux-coordinate
independent approach within the discontinuous Galerkin framework~\cite{Held2016}, which we are the first to apply to full-F gyro-fluid models~\cite{WiesenbergerPhD, HeldPhD}.
Recent studies focus on numerical elliptic grid generation~\cite{wiesenberger2017,Wiesenberger2018}. Both are important for the efficient description of realistic magnetic field geometries.

One of the main features of the code are matrix (and in general container) free algorithms. This type of algorithm ignores
the exact format or implementation of the matrix (or vector) type employed.
In consequence a matrix-free implementation offers a highly flexible framework with respect to both the equations discretized and the hardware the code runs on.
It allows the development of platform independent code, with the compiler
choosing
implementations for Nvidia GPUs using the CUDA programming language,
the OpenMP parallelized version for CPUs~\cite{einkemmer2014} or Xeon Phi co-processors, or
the immediate extension to
hybrid parallelization using the message passing interface (MPI).

% CPC Topics: (Papers describing (...) established software of importance in the community are encouraged)
% - Contemporary computational methods and techniques and their implementation. The effectiveness in the context of a substantial problem in physics.
%
% - Compuational models and programs
% - mathematical and numerical methods and algorithms
% - the impact of advanced computer architecture and special purpose computers on computing in the physical sciences
% - software topics related to, and of importance in, the physical sciences

In Section~\ref{sec:feltor} of this article we give a short overview over the structure
and goals of the \feltor project.
Then, in the following two sections we focus on reproducibility, accuracy and performance of the library.
%Please note that in order to ease the reading
%of the article we moved the introduction to each of the discussed topics to the
%respective Section.
In Section~\ref{sec:reproducibility} we show how round-off errors
caused by the machine precision can destroy accuracy and reproducibility of a simulation.
We demonstrate the implementation steps necessary
to restore accuracy and \textit{bitwise} reproducibility and then debate in what ways a simulation of
an ill-conditioned set of equations can be reproducible.
%In Section~\ref{sec:arakawa} we present a qualitative comparison of the Arakawa algorithm
%with its discontinuous Galerkin version at the example of Euler's equation in polar
%coordinates.
In Section~\ref{sec:performance} we present results of a performance study.
We briefly discuss important performance tuning methods and
derive a parallel model that can predict the runtime of any algorithm in \textsc{feltor} on a variety of computer architectures.
Finally, we present an overall discussion and conclusion of our results in Section~\ref{sec:conclusion}.
%
%Each of these projects follows different goals and is in principle independent
%from the others.
%However, through the integration in \feltor synergies, but also conflicts of
%interest between the various stakeholders, appear, which we critically discuss in the Conclusion.
%

%% file: feltor.tex
\section{\feltor overview} \label{sec:feltor}
In this Section we give a brief overview of the structure
of the \feltor project and outline its design goals and motivation.
Furthermore, we shortly discuss the most
important implications of the project structure and conclude the 
Section with
a small code sample as a first impression of the library usage.
%code will not be discussed

The details of how we realize the structure and our design
goals in code are absent in this discussion but are available
in the accompanying code repository~\cite{Feltor-v5.0}.
In general, we use design principles similarly found in other
existing code projects (for example \cite{cusp}) and
as far as possible try to adhere to
established coding practices \cite{meyers2005, meyers2014,alexandrescu2001}.
%documentation provides more details to the discussion
%The code repository
%includes instructions on how to compile the full user
%documentation,
%which is also available on our homepage~\cite{Homepage}.
%
Furthermore, we invite the interested reader to explore
our homepage~\cite{Homepage} in parallel to the current section for additional information and details.

\subsection{Overview}
\begin{figure}[htpb]
  \centering
  \includegraphics[width=0.48\textwidth]{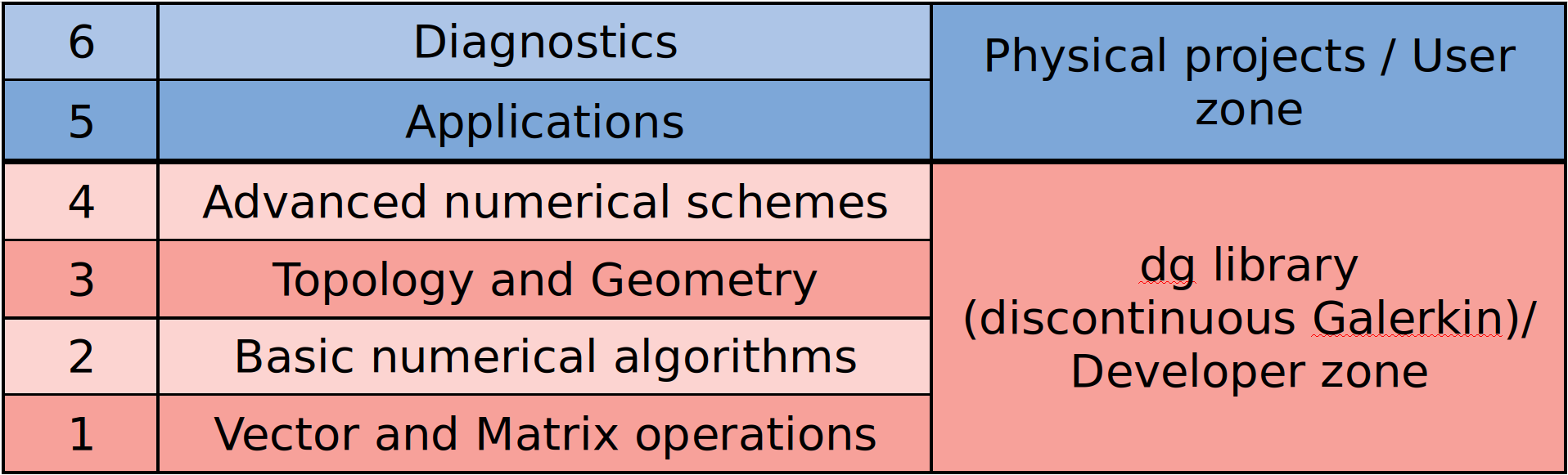}
  \caption{The structure of the project:
  \feltor is both a numerical library and a scientific software package built on top of that library.
}
  \label{fig:design}
\end{figure}

\feltor (Full-F \textsc{EL}ectromagnetic code in {\textsc{TOR}}oidal geometry) is a modular
scientific software package that can be divided into six layers.
Each layer defines and implements an interface that can be used by the same
or higher levels. This structure is depicted in Fig.~\ref{fig:design}. In the following we shortly introduce each layer and the 
capabilities it adds to the library.
%%%%%%%%%%%%%%%%%%%%%%%%%%%%%%%%%%%%%%%%%%%%%%%%%%%
%%%%%%%%%%%%%%%%%%%%%%%%%%%%%%%%%%%%%%%%%%%%%%%%%%%
 \begin{description}
 \item[User Zone] A collection of actual {\color{blue} simulation projects and diagnostic
programs} for two- and three-dimensional drift- and gyro-fluid models
\begin{description}
  \item [6 Diagnostics]
    These programs are designed to analyse the output from the application programs.
  \item [5 Applications]
    Programs that execute two- and three-dimensional simulations: read in input file(s), simulate, and either write results to disc or directly visualize them on screen. Some examples led to 
journal publications in the past \cite{wiesenberger2014,Held2016a,Kube2016, wiesenberger2017b}.
\end{description}
\item [Developper Zone] The core {\color{purple} dg library} of optimized (mostly linear algebra) numerical algorithms
and functions centered around discontinuous Galerkin methods on structured grids. Can be used as a standalone library.
\begin{description}
  \item [4 Advanced algorithms]
    Numerical schemes that are based on the existence of a geometry and/or a topology. These include for example the discretization of elliptic equations in arbitrary coordinates, multigrid algorithms and
a semi-Lagrangian scheme to compute directional derivatives
along arbitrary vector fields~\cite{Held2016}.
  \item[3 Topology and Geometry]
    Here, we introduce data structures and functions that represent the concepts of Topology and Geometry and operations defined on them (for example the discontinuous Galerkin discretization of derivatives~\cite{einkemmer2014}).  The {\it geometries} extension implements a large variety of grids and grid generation algorithms that can be used here~\cite{wiesenberger2017,Wiesenberger2018}.
  \item[2 Basic algorithms]
    Algorithms like conjugate gradient (CG) or Runge-Kutta schemes that can be implemented with basic linear algebra functions alone.
  \item[1 Vector and Matrix operations]
    In this "hardware abstraction" level we define the interface for a set of various vector and matrix operations like additions, multiplications, and scalar products. These functions are implemented  and optimized on a variety of hardware architectures
and serve as building blocks for all higher level algorithms. We study those in Section~\ref{sec:performance} of this contribution.
\end{description}
\end{description}
\subsection{Implications of the code structure}
%work on each level influences the others
It is possible for several groups to work independently
on and with \textsc{Feltor} on the various levels outlined in Fig.~\ref{fig:design}.
Combining the defined building blocks from the lower levels a user can
freely construct and explore new numerical algorithms or physical equations.
At the same time
any improvement or upgrade of the core level routines
improves the performance of all application codes using it.
Of course, the set of primitive functions also restricts the number of possible
numerical algorithms or equations that can be implemented.
For example direct solvers are absent in \textsc{Feltor}.
%An application is therefore also a driving force behind extending the library
%with new features at lower levels.

Another advantage is the possibility to test functions and modules separately
and independently of each other. We use this feature extensively throughout the
development process on all levels outlined in Fig.~\ref{fig:design}.
Specifically, our tests encompass unit tests for low level subroutines,
convergence studies of specific numerical algorithms
as well as conservation studies of invariants in our physical models.

%stakeholders don't always share interest
\subsection{Design goals}
The implementation of \feltor is the result of an
ongoing development process and subject to frequent changes.
In the following we thus rather describe our goals and guidelines.
These have led to the present state of the code and will likely prevail in the future.
\begin{description}
    \item[Code readability]
      Numerical algorithms can be formulated clearly and concisely. In particular, parallelization strategies or optimization details are absent in application codes.
    \item[Ease of use]
      We try to make our interfaces as intuitive and simple as possible. It is possible for \CC beginners to write useful, fast and reliable code with \textsc{Feltor.}
        This feature is enhanced by an exhaustive documentation and tutorials on our homepage~\cite{Homepage}.
    \item[Fast development] 
      A particular important feature from the user perspective
is the possibility to quickly set up or change model equations
in a minimum amount of time. We accomplish this feature by 
providing building blocks at \textsc{Feltor}'s core levels, which
can be freely combined or rearranged. 
    \item[Speed] 
      \feltor provides specialized versions of the performance critical Level 1 functions for various target hardware architectures including for example GPUs and Intel Xeon Phis. 
Note that writing parallelized code is the default in \textsc{Feltor.} We explore and discuss performance critical issues in Section \ref{sec:performance} of this article.
    \item[ Platform independent] 
      Application code runs unchanged on a large variety of hardware ranging from a desktop environment to mid-sized compute clusters with 
dedicated accelerator cards. The library adapts to the resources 
present in the system and chooses correct implementation of functions at compile time. 
This is possible through a template traits dispatch
system in combination with classic C-style macros at \textsc{Feltor}'s core level.
We demonstrate this feature explicitly in Section \ref{sec:performance}
of this article.
    \item[ Extensibility] 
      The library is open for extensions to future hardware, new numerical algorithms and physical model equations.
    \item[ Defined scope] 
      Our focus lies on efficient discontinuous Galerkin methods on structured grids and their application to drift- and gyro-fluid equations in two and three dimensions. 
      We outsource any other operation, in particular input/output, to external libraries.
\end{description}

\subsection{Getting started}
\begin{figure}[htpb]
  \centering
\includegraphics[scale=0.6]{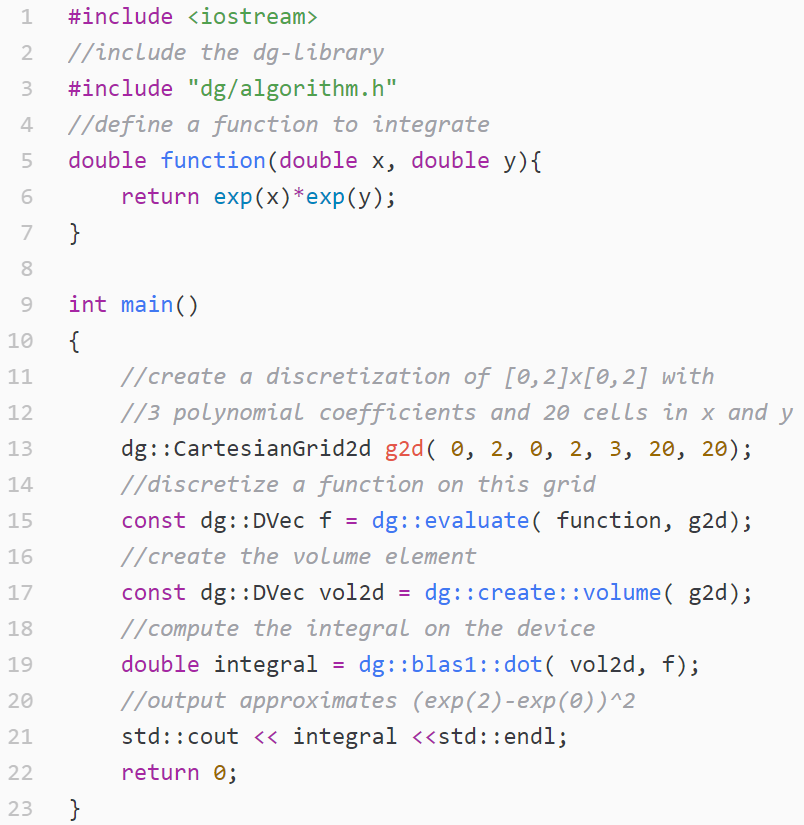}
    \caption{ Sample application using \textsc{Feltor}. For details on the function calls
    see the online documentation. 
 }
    \label{fig:code_sample}
\end{figure}

We provide a "Quick start guide" contained
in the README file of the dataset~\cite{Feltor-v5.0}, which explains how to setup
our library on various systems.
In Fig.~\ref{fig:code_sample}, we show a small teaser program to give
readers a first impression how code using \feltor looks like. It
integrates the function $f(x,y) = \exp(x) \exp(y)$ on the domain 
$[0,2]\times[0,2]$ using Gauss-Legendre integration.
Depending on how this program is compiled the main computation of 
the scalar product in line 19
executes either on a GPU or on a shared memory host system.
In fact, line 19 is the first example of platform independent code. 
The \texttt{dg::blas1::dot} function is a template that chooses the 
implementation based on the vector class it is called with. 
This means that we could also generate an MPI grid in line 13 and change
the \texttt{dg::DVec} to \texttt{dg::MDVec} (an MPI distributed device vector).
Then, \texttt{dg::blas1::dot} reroutes to the MPI implementation of the scalar product function. Notice that \texttt{dg::blas1::dot} implements the exact algorithm
we discuss in Section~\ref{sec:reproducibility}.

Unfortunately, a more detailed description of the library surpasses the scope
of the present article. However, the interested reader will find a helpful tutorial
on our webpage\cite{Homepage}, which gives a step-by-step introduction 
to the library and shows and explains many practical code examples. 
As mentioned earlier, the webpage also contains a more formal documentation of all functions and
classes the library provides and pdf files that describe our numerical methods.
Hopefully, this will convince the reader that we achieve the design goals outlined previously.

%% file: reproducibility.tex
\section{Reproducibility in numerical simulations} \label{sec:reproducibility}
A paradigmatic model to study drift wave turbulence and zonal flow dynamics in the edge of magnetized fusion plasmas is
the Hasegawa-Wakatani (HW) model~\cite{hasegawa83,wakatani84,wakatani87,numata07}. 
Recently, this model has been extended to include large relative density fluctuation amplitudes and steep density gradients within a full-F gyro-fluid approach, 
thus facilitating studies in the non-Oberbeck-Boussinesq regime~\cite{Held2018}. 
The dimensionless modified full-F HW equations consists of continuity equations for electron particle density \(n\), ion gyro-center density \(N\) and the polarisation equation
\begin{subequations}
\begin{align}
\label{eq:fullFdtne}
 \partial_t n  + \left\{ \phi,n \right\} &=\alpha \left(\widetilde{\phi}-\widetilde{\ln\left(n\right)} \right) ,\\
\label{eq:fullFdtNi}
\partial_t N   + \left\{ \phi -({\nabla} \phi)^2 / 2,N \right\}  &=0 , \\
 {\nabla} \cdot \left(N {\nabla}_\perp \phi\right)&= n- N, 
\end{align}
\label{eq:hw}
\end{subequations}
with electric potential \(\phi\), adiabaticity parameter \(\alpha\) and Poisson bracket $\{f,g\} := \partial_x f \partial_y g - \partial_y f \partial_x g$.
The Reynolds decomposition \(f := \langle f  \rangle + \widetilde{f}\) with Reynolds averaged part \(\langle{  f}\rangle := L_y^{-1}\int_0^{L_y} dy \hspace{1mm} f\) 
and fluctuating part \(\widetilde{f}\) is utilized in the parallel coupling term on the right hand side of Eq.~\eqref{eq:fullFdtne}.

The initial (gyro-center) density fields \(n(\vec{x},0)=N(\vec{x},0) = n_G(x)\left(1+\delta{n}_0(\vec{x})\right)\) consist of the reference background density profile \(n_G:=e^{-\kappa x}\), 
which is perturbed by a turbulent bath \(\delta {n}_0(\vec{x})\). Here, \(\kappa\) parameterizes the constant background density gradient length. For further details to the model we refer the reader to Reference~\cite{Held2018}.

\begin{figure}[htpb]
  \centering
\includegraphics[scale=0.45]{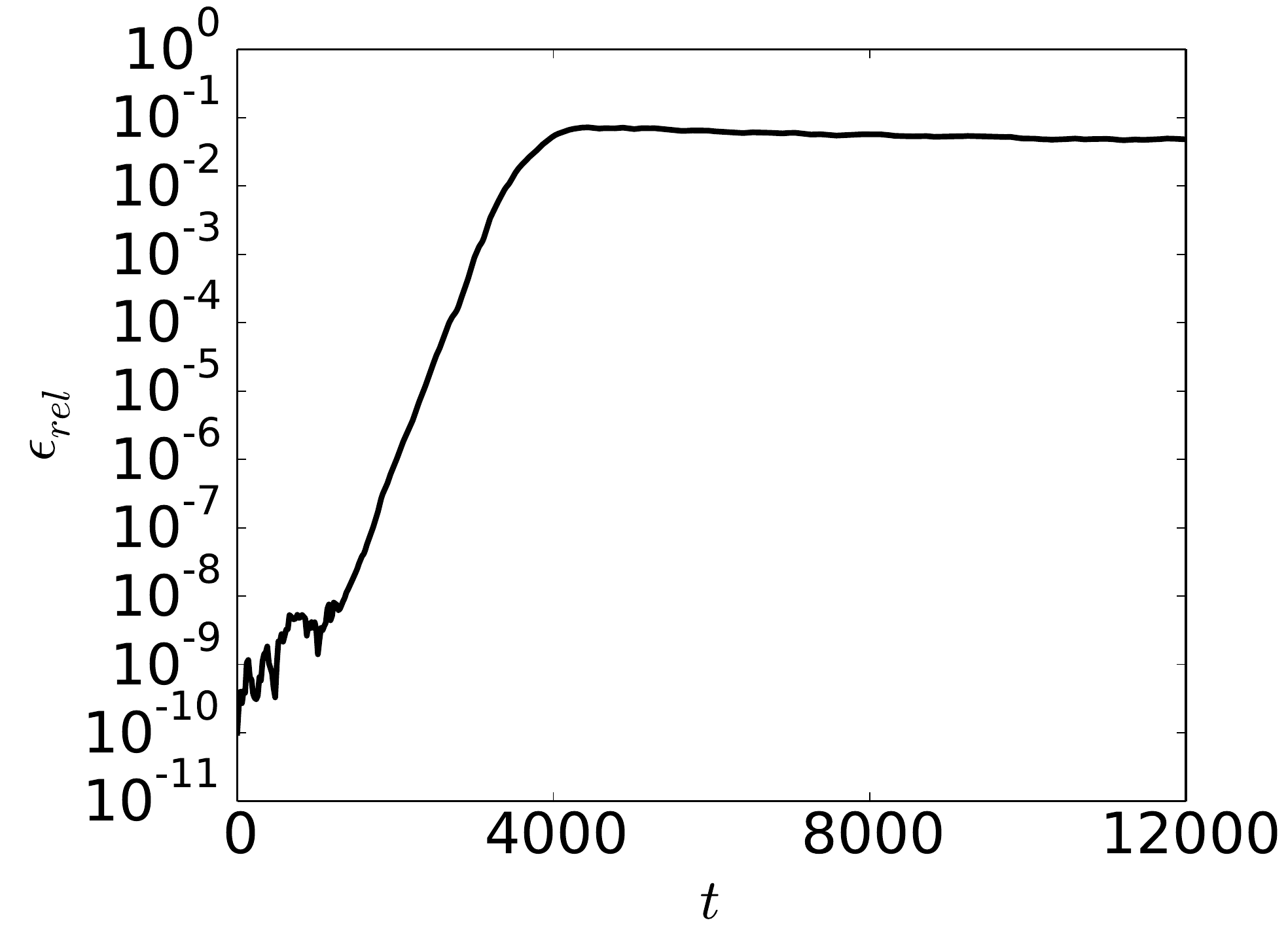}
\caption{The relative error \(\epsilon_{rel}\) as a function of time \(t\) is depicted. The relative error \(\epsilon_{rel}\) between the na\"ive runs increases towards \(\mathcal{O}(10^{-1}) \). Note, that the relative error \(\epsilon_{rel}\) is biased by the constant \(10^{-10}\).
 }
    \label{fig:compare}
\end{figure}

We implemented Eqs.~\ref{eq:hw} in \feltor and now want to test the
reproducibility of our parallel simulations. More precisely, we want to test
if with the exact same input parameters our executable reproduces the exact same
output in subsequent runs.
To this end, we fix a typical set of physical and numerical input parameters (contained in the repository~\cite{Feltor-v5.0}) %, which consist of an
%inverse background density gradient length \(\kappa=1/32\), adiabaticity \(\alpha=0.001\) and box size \(L_x=L_y=128\). %MW the interested reader can look these up in the dataset
and run our executable twice with the exact same initial condition and parallelization strategy.
In Fig.~\ref{fig:compare} we compare the output of the two runs at each time step.
Initially, the relative error \(\epsilon_{rel}:= || n_1 - n_2||_2/||n_1 ||_2\) between the two solutions vanishes.
Here, \(n_1\) and \(n_2\) is the electron density of the first and second simulation, respectively, and \(|| f||_2\) is the \(L_2\) norm.
As time advances $\epsilon_{rel}$ rapidly increases towards \(\mathcal{O}(10^{-1}) \).

 Although this result is very surprising at first, the possibility for two identical simulation setups to have non-identical results is
 readily explained. First, let us recall the finite nature (64-, 32-, or 16-bits) of floating-point
 computations that results in the non-associativity of floating-point operations~\cite{Goldberg1991}.
For instance, let us denote $\oplus$ as the addition in \binary\ floating-point arithmetic, then
$(-1 \oplus 1) \oplus
2^{-53} \neq -1 \oplus (1 \oplus 2^{-53})$ since $(-1 \oplus 1) \oplus
2^{-53} = 2^{-53}$ and $-1 \oplus (1 \oplus
2^{-53} )=0$.
Second, in a parallel environment the order of execution between threads is usually arbitrary
and can vary between runs.
Therefore, subsequent runs of a parallelized executable with identical input may indeed produce various binary outputs.
On the other side, if the small round-off errors of machine precision lead to
a large error in subsequent simulation times as seen in Fig.~\ref{fig:compare},
then we are apparently faced with an ill-conditioned problem.

Both the fact that executables may produce non-deterministic results and the fact that
small derivations may grow exponentially in ill-conditioned problems
raise concerns about our ability to reproduce and verify our numerical simulations.
In the following we view these concerns from various angles.
First, we discuss reproducibility and accuracy
from a purely computational and programmatic viewpoint.
In Sections~\ref{sec:bitwise} and \ref{sec:programmatic} we present how
with the help of so-called \textit{long accumulators} together with \textit{floating-point expansions} and \textit{error-free transformations} 
we can achieve \textit{bitwise reproducibility} in our simulations.
In the following Section~\ref{sec:physical}, we then view the problem from a broader perspective and take computational,
numerical and physical considerations into account.
We debate the implications of finite machine precision and ill-conditioned
problems on the accuracy, convergence, reproducibility, and verification of numerical simulations.

\subsection{ExBLAS: Accurate and bitwise reproducible Basic Linear Algebra Subprograms (BLAS)} \label{sec:bitwise}
%MW: first paragraph split and moved into introduction/discussion
%Floating-point arithmetic consists in an approximating real numbers with a significand, an exponent, and a sign:
%\[
%x = \pm \underbrace{x_0 . x_1 \ldots x_{M-1}}_{mantissa} \times b^{e}, \quad 0 \leq x_i \leq b-1, \quad x_0 \neq 0,
%\]
%where $b$ is the  basis ($2$ in our case), $M$ is the precision, and $e$ stands for the exponent that is bounded 
%($e_{\min} \leq e \leq e_{\max}$).

In this article, we consider the \binary\ or double-precision format of the
IEEE-754-2008 standard, which requires the basic arithmetic operations $(+, -,
\times , /, \sqrt{~})$ to be correctly rounded
(rounding-to-nearest)~\cite{Goldberg1991,Higham02,HFPA10}.
Thanks to the fact that most processors implement this standard, the numerical portability of applications was eased. 
Due to the finite nature of floating-point computations as well as the non-determinism of parallel executions,
we develop an approach to ensure bit-wise reproducibility via ensuring correctly rounded results, whenever possible. 
The main idea is to keep track of both the result and the errors during the course of computations.
To increase the accuracy of floating-point operations, i.e. assure their correct rounding, we rely upon the 
following two strategies: the first computes the result and recovers the rounding error using so-called \textit{error-free transformations} (EFT)
and stores both result and error in a \textit{floating-point expansion} (FPE). A FPE is an unevaluated sum of $p$ floating-point numbers whose 
components are ordered in magnitude with minimal overlap to cover a wide range of exponents. 
Typically, a FPE relies upon the use of the \twosum\ EFT~\cite{Knu97} for the addition and the use of the \twoprod\ EFT for the 
multiplication~\cite{Ogita05accuratesum}. 
%\twosum\ computes the addition $r$ of two floating-point numbers $a$ and $b$ and 
%the rounding error $e$ such that $r$ and $e$ do not overlap and $a + b = r + e$. 
%Similarly, \twoprod\ computes the product $r$ of two floating-point numbers $a$ and $b$ as well as the rounding error $e$. For \twoprod, we use the fused-multiply-and-add (\fma) instruction to track the error that  
%computes $a \cdot b - r$ with only one rounding at the end.
The main advantage of FPEs is that they could be fetched to the registers and reside there during the computation.
However, they may not be able to guard every bit of information, which is necessary for correct rounding, for large sums 
or for floating-point numbers with significant variations in magnitude.

The second strategy projects the finite range of exponents of floating-point numbers into a long vector the so-called 
\textit{ long (fixed-point) accumulator}. 
A fixed-point representation stores numbers using an integral part and a fractional part of fixed size, or equivalently a scaled integer.  For instance, Kulisch~\cite{Kulisch11} proposed to use a 4288-bit long accumulator for the exact dot product of two vectors composed of \binary\ numbers; however, such a large long accumulator is designed to cover 
all the severe cases without overflows in its highest digit. By preserving every bit of information, the long accumulator guarantees 
to compute the exact result of a large amount of floating-point numbers of arbitrary magnitude. However, when comparing to FPE,
the long accumulator has a large memory footprint and requires roughly two times more operations to be performed.

With the aim to derive fast, accurate, and reproducible Basic Linear Algebra Subprograms (BLAS), we construct 
a multi-level approach for these operations that is tailored for various modern architectures with their
complex multi-level memory structures.
On one side, we want this approach to be fast to ensure similar performance compared to the non-deterministic parallel
versions. On the other side, we want to preserve every bit of information before the final rounding to the desired format to assure
correct-rounding and, therefore, reproducibility. To accomplish our goal, we merge together FPE and long accumulators, tune them,
and efficiently implement them on various architectures, including conventional CPUs, Nvidia and AMD GPUs, and Intel Xeon Phi co-processors (for details we refer to Reference~\cite{Collange15Parco}).

We begin with the parallel reduction, which is at the core of many BLAS routines. We build its scalable, accurate, 
and reproducible version using FPEs with the \twosum\ EFT~\cite{Knu97} and long accumulators. In practice, the latter is so rarely invoked that
only little overhead (less than $8$\,\%) results on summing large vectors.
The dot product of two vectors is another crucial fundamental \blas\ operation.
The \exdotp\ (exact stands for accurate and reproducible) algorithm is based on the previous \exsum\ algorithm and 
the \twoprod\ EFT~\cite{Ogita05accuratesum}: we accumulate both the result and the error to FPEs and reduce these FPEs and long accumulators on various levels as in \exsum.
These and other routines -- such as matrix-vector product (\exgemv), triangular solve (\extrsv), and 
matrix-matrix multiplication (\exgemm) -- are distributed as the Exact BLAS (ExBLAS) library~\cite{Iakymchuk15ExBLAS,exblas}.
Thanks to the modular and hierarchical structure of linear algebra algorithms, higher level operations -- such as matrix factorizations 
-- can be entirely built on top of the fundamental kernels as those in the BLAS library. In ExBLAS, we follow this principal to construct reproducible LU factorizations with partial pivoting.
%%%%MW:
\subsection{Reproducibility in \feltor}\label{sec:programmatic}
As outlined in Section~\ref{sec:feltor} \feltor builds its algorithms on basic
primitive functions, which partly overlap with the BLAS library.
Please find the exact list of functions in the documentation.
Our basic assumption is that, if these elementary functions are reproducible,
then all algorithms and simulations implemented with them are reproducible. This assumption
follows our theoretical and practical studies~\cite{iakymchuk:hal-01419813} of the unblocked LU factorization with partial pivoting,
which underneath is entirely build upon the BLAS routines.
The first step to realize our goal
incorporates the correctly rounded and reproducible parallel reduction from the ExBLAS library into \textsc{Feltor}.
In this way, we can provide the accurate and reproducible dot product $\sum_i x_i y_i$.
Note that we also provide a function computing the weighted
sum $\sum_i x_iw_i y_i$, where $w$ represents for example the volume form of our coordinate system.
This is important in numerical computations of the scalar product $\int f_1 f_2\sqrt{g} \text{d}V$ with functions $f_1$, $f_2$ and volume element $\sqrt{g}$.

In the second step we make the trivially parallel vector operations like $y\leftarrow \alpha x+\beta y$ reproducible.
Unfortunately, the use of FPEs or long accumulators for these very small summations
introduce too much overhead to be practical.
Hence, our idea is to just fix the type and order of execution of floating point operations. The results should then be identical for all compilers and platforms
that follow the IEEE-754-2008 standard.
% Hence, we combine the ExBLAS approach with the careful arrangement of computations in such a way that the non-determinism is eliminated. % see the $ab + cd$ example below. %\fixme{On the other hand we do not parallelize the summation itself -- not clear too me, RI. The next sentence also needs to be fixed} 
%We therefore use that if we can guarantee the type and order of execution to be the same
%on all compilers and platforms that follow the IEEE-754-2008 standard, the results are identical
%even though they are not correctly rounded.
The problem with this is that for performance reasons,
the \CC language standard allows compilers to change the execution order of a given
line of code. It even allows merging multiplications and summations with fused multiply add (FMA) instructions. This computes \texttt{a*x+b} in a single instruction
with only a single rounding operation at the end.
Let us consider the 'naive' implementation \texttt{y=a*x+b*y}.
A compiler might translate this to two multiplications \texttt{t1=a*x} and \texttt{t2=b*y}
and a subsequent summation \texttt{y=t1+t2}; it might generate a single multiplication \texttt{t=b*y}
with a subsequent FMA \texttt{y=fma(a,x,t)}, which gives a slightly different result; or it may even compute \texttt{t=a*x} first and then use the FMA \texttt{y=fma(b,y,t)}.

Our approach to solve this issue is to explicitly instruct the compiler to use FMAs
together with relevant compiler flags to prevent the use of value changing optimization techniques (for example \texttt{-fp-model precise} for the Intel \texttt{icc} compiler).
The former is possible through the \texttt{std::fma} instruction added to the
\CC-11 language standard\footnote{Unfortunately, at the time of this writing
  the Intel and Microsoft compilers do not properly vectorize code involving \texttt{std::fma}.
  For the time being our implementation relies on \texttt{icc} and \texttt{msvc} to
  always translate \texttt{a*x+b} into an FMA instruction.}.
With this combination we avoid non-determinism in the order of operations, reduce the number of rounding errors from three to two, and, therefore, achieve bitwise reproducibility for this operations and even for matrix-vector
multiplications $y\leftarrow \alpha Mx+\beta y$. Again, we do not use long
accumulators for the summation but only fix the order of execution.
However, we need to take special care
in our MPI implementation. The computation of
boundary points can begin only after all values from other processes were
communicated.

The third step towards reproducibility in \feltor is to make the initialization of
vectors reproducible. Here, the main problem lies in the use of transcendental functions like $e^x$,
$\sin(x)$ or $\cos(x)$. % Consider for example Eq.~\eqref{eq:initial} in Section~\ref{sec:arakawa}.
 The algorithms for computing these functions differ by compiler
and the results subsequently differ if not correctly rounded\footnote{In fact, the difference comes from the transcendental functions implementations in \texttt{libm}. Note that GNU \texttt{libm} ensures correct-rounding of these functions thanks to the GNU Multi Precision Arithmetic library. With \texttt{icc} we had to use a special flag \texttt{-fimf-arch-consistency=true} to
get reproducible results across platforms.}. A practical and portable solution to this
problem is an open issue in \textsc{Feltor}.

All in all, \feltor yields reproducible results up to the compiler and the hardware's
capability to compute FMAs.
This means that we can reproduce simulation results bit for bit, independently
of parallelization, as long as we
use the same compiler and the CPU is capable of computing FMAs.

\subsection{Bitwise reproducibility, accuracy, convergence and verification}\label{sec:physical}
\begin{figure}[htpb]
  \centering
\includegraphics[scale=0.45]{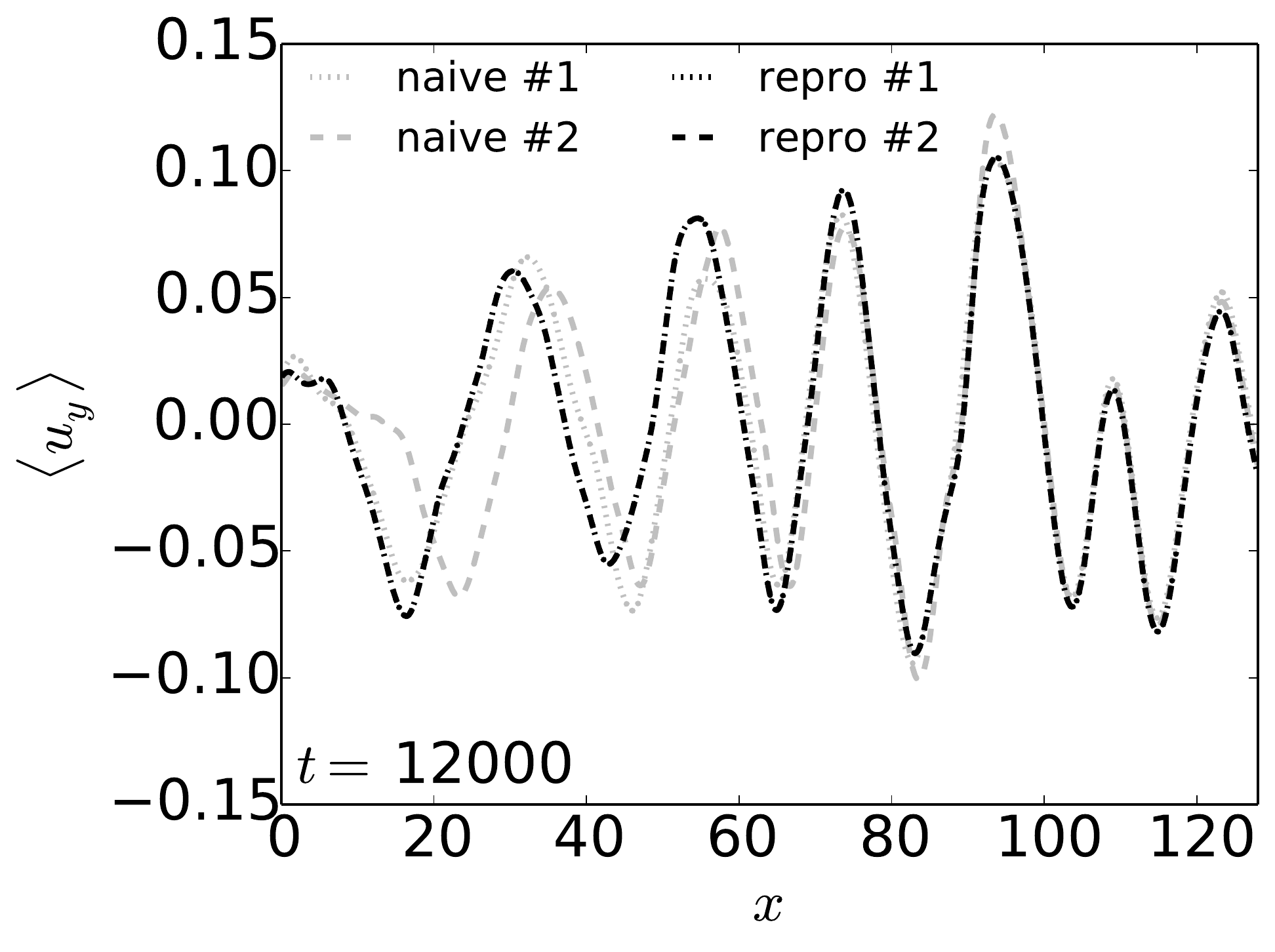}
  \caption{The radial zonal flow signature is shown. The deviation of zonal flow structure of two na\"ive runs with identical initial conditions is clearly visible. As opposed to this the
  zonal flow structure of the bitwise reproducible runs are identical.}
    \label{fig:zf}
\end{figure}

    We improved \feltor with the reproducible BLAS Level-1 subroutines and can now
    re-simulate Eqs.~\ref{eq:hw}
    and indeed obtain bitwise identical results after each run. Please
    consult the dataset~\cite{performance} introduced in Section~\ref{sec:performance} for details on the exact compiler
    flags and hardware that we use for our simulations.
    We show our solution in Fig.~\ref{fig:zf} where we compare the
    radial zonal flow structure to the previous implementation.
    Here, the radial zonal flow structure of the na\"ive implementation deviates while the zonal flow structures are identical in the new (reproducible) implementation.

Let us summarize the implications of what we have achieved and know up to this point.

\begin{description}
  \item[Bitwise reproducibility]
    We have the possibility to reproduce parallel simulation results bit-to-bit.
    This is particularly advantageous from a programmatic
    point of view since alterations in the implementation or future
    adaptions to other parallel hardware can be rigorously checked and tested.
    Moreover, independent outside groups and ourselves gain the possibility to
    re-simulate and confirm the results. This is especially important since
    we usually refrain from publishing output files due to their impractically large size.
    Now, we have the possibility to publish the code together with input files
    and can expect to exactly reproduce presented results~\cite{Feltor-v5.0}.
    %\fixme{Have you also observed that the number of iterations in CG and the error are identical while using the reproducible implementation which is not the case for the standard one?} %MW: of course, else it wouldn't be reproducible?
  \item[Accuracy]
    It is important to mention that we not only achieved bitwise reproducibility
    but also increased the accuracy of our implementation, mainly in the scalar product. The latter is for example important in a conjugate gradient method.
    The problem with the previous na\"ive summation was the unfavourable cancellation
    of digits when adding small values to a large sum.
    Even in a tree summation
    algorithm the error grows with the size of the array with $\sqrt{a}$ with $a$ being
    the array size.
        This effectively reduces the machine precision, which is a particular concern for large scale single precision computations. It is expected that the next generation of supercomputers -- such as Exascale systems delivering $10^{18}$ floating-point operations per second -- will be composed of heterogeneous resources like CPUs and accelerators. Obtaining peak performance on these systems will, in all likelihood, require the use of single or mixed-precision simulations \cite{ascac,baboulin2009,chen2012,einkemmer2016mixed}.
    %In order to efficiently utilize such
  %resources, runtimes would have to employ sophisticated strategies for dynamic data management and task
  %scheduling. Such optimization approaches will obviously be beneficial for performance, but may violate
  %the accuracy (and reproducibility) of simulations.

  \item[Condition]
      If we evolve the physical model Eqs.~\ref{eq:hw} over long periods of time, even small (physical) perturbations in the initial state can be amplified by many orders of magnitude.
      This is a fundamental property of the physical system under consideration. 
      Consequently, this behaviour is also reflected in the numerically discretized system of equations.
      Recall that (numerical) perturbations
      are always present in this system.
      For example, even if the initial state is given by an
      analytical function its numerical representation is already inexact due to either the discretization error or the finite precision of floating point arithmetics. These (numerical) perturbations then grow over time just as their physical counterparts do.

    %After linearization and discretisation of Eqs.~\ref{eq:hw} it is evident that the discrete linear system is in general non-normal and consequently the eigenvectors are non-orthogonal.
%As a result the condition number of the linear system is for realistic parameters very large and any perturbation may be amplified by the condition number.
%In particular for a linear system \(\frac{||\delta \vec{x} ||}{|| \vec{x} ||}\leq || A ||  || A^{-1} || \frac{||\delta A ||} {|| A ||}\) \fixme{elaborate a bit more}.
%The condition thus indicates how fast two very close initial states will drift apart.
\end{description}

%Let us now discuss overall the implications of the here presented points.
In conclusion, we have to accept that even with the increased accuracy and reproducibility of our implementation
the error\footnote{ in the $L^\infty$ or any other suitable norm}
in our numerical solution is large after a sufficiently long simulation time.
This is because any error stemming either from the numerical discretization
or the finite machine precision will be amplified by the system.
In particular this means that we cannot obtain convergence of our simulation.
Even with infinite machine precision
we would need a prohibitively fine grid to find the exact solution.
On the other side, the error in a single time-step or
small enough time span may still be acceptable and converge with the
expected order. For example in Fig.~\ref{fig:compare} the error
from machine precision is still small for times below $t=2000$ say, such
that a convergence study is possible in this regime.
In this context, we can also expect that as long as we can maintain the machine
precision in our implementation (especially for the dot product, as discussed above)
using single precision gives the same physical results as does double precision. In memory bandwidth bound problems this can potentially lead to a factor two gain in performance.

Furthermore, we have to reject the notion that our bitwise reproducible solution
is any more physically or numerically reasonable than the previous solution, even
if the accuracy of elemental operations was increased.
We only select one specific solution out of a larger class of solutions equivalent
within the limits of the accuracy of the numerical discretization
and the machine precision.
In fact, we can also physically expect a larger class of end states that
are equivalent within small (thermal) fluctuations that are present in the turbulent system described by Eqs.~\ref{eq:hw}.

All of these points raise the question: how and in what sense can we then
consider simulation results to be "correct"?
%We therefore conclude that we need alternative methods to verify
%our numerical representation of Eqs.~\ref{eq:hw} than pointwise convergence.
The answer to this question depends on what the goal of the
simulation effort is.
A first step is to identify quantities that are of physical interest. This can for example be the zonal flow structure in Fig.~\ref{fig:zf} or turbulent spectra. Convergence can then be studied directly in terms of the physical quantities that we are interested in. In many situations, these quantities show convergence even though pointwise convergence of the numerical solution can not be achieved (see, for example, \cite{einkemmer2018perfcomp}).
In addition, we recommend checking that the quantities of interest are insensitive to very small perturbations. This can be done by performing simulations of an ensemble of slightly perturbed initial values. If the quantities of interest have similar values for the entire ensemble, confidence in the corresponding numerical results is increased.
%\fixme{this is precisely backward error according to Wilkinson and Higham. This refers to a method is backward stable if $y = f(x+\delta x)$ where $\delta x$ is very small. I need to explore it further. Lukas may comment it better from numerical analysis point of view with more appropriate words. If that does not hold, often, eg LU, some strategies are appleid like pivoting. Is this possible with your method? But that is for ill-conditioned problems or close to that}.

Finally, invariants of the physical model can be used as
a consistency check of the numerical method and the corresponding implementation.
Also, they help to restrict the range of possible solutions to the system.
This, on one hand, means that we should favour physical models that do provide
invariants and numerical methods that conserve these invariants.
Conservative numerical methods can (often) give us a good physical picture even though the $L^\infty$ error is large.
On the other hand, it is difficult to guarantee correct (physical) behaviour from symmetries of a system alone.
For example energy conservation is in general not enough to guarantee physical
solutions. As exemplified by the integration of the solar system in \cite[Chap. 1]{hairer2006geometric},
a numerical integrator may be locally converged and conserve energy
and still produce a strikingly wrong (both qualitative and quantitative)
result after a long time.
%
%Thus, we still have to conduct convergence studies applied to the physical quantities of interest.
Ultimately, we will have to accept that if the physical interest \textit{is} the
exact result (for example the precise velocity fluctuations in a turbulence simulation), the possible simulation time span
is largely reduced depending on the condition of the system. We note again that
this is not an algorithmic or implementation problem. It is rather a consequence of the physical properties of the system that we investigate.

%%
%Thus, various optimization strategies, data partitioning
%patterns, reduction trees, thread/grid shapes potentially impact the computed results since they change the order of the 
%underlying floating-point operations.
%%

%% file: performance.tex
\section{Performance and runtime prediction} \label{sec:performance}

%In the last years, power consumption in High Performance Computing (HPC) systems have become one of the limiting characteristics in the road map for the realization of Exascale computing. Therefore, energy efficiency turned into a necessary trend in the design criteria for emerging HCP hardware. Architecture with lower clock frequency and more powerful vectorization units have aroused. These new architectures coexist with multicores which have been also increasing the number of cores, but relying in higher clock frequency. Both architectures have in common that the exploitation of the potential performance is responsibility of the developer and in this section we will expore the techniques to achieve it.

In this Section we present a performance study of the \feltor library. This study includes a second dataset~\cite{performance} to this article, which provides the complete raw data in csv format as well as the \texttt{ipython} notebooks used for the data analysis and plot generation. 
%The notebooks can be viewed online \url{https://github.com/mwiesenberger/performance}.
The interested reader is invited to inspect these notebooks in parallel to reading this section for additional information and details.

We begin this section with a discussion of important performance optimization techniques for memory bandwidth bound algorithms~\ref{sec:optimization}. We then shortly describe the hardware, the configuration and the program that we used to generate the performance data~\ref{sec:configuration}.
Having measured and discussed the performance of \textsc{Feltor's} building  blocks in Section~\ref{sec:measurements} we suggest a performance model that predicts the runtime of any constructed algorithm in Section~\ref{sec:prediction}.
%We show that with the exception of the multi-node knights landing architecture and a system with a very fast cache the predicted runtime matches the measured runtime up to $25\%$ for a large variety of hardware architectures and problem sizes.
We discuss strong and weak scaling in Section~\ref{sec:scaling} and conclude with
a critical discussion in Section~\ref{sec:performance_discussion}.

%%%%%%%%%%%%%%%%%%Albert and Xavier%%%%%%%%%%%%%%%%%%%%%%%%%%%%%%
\subsection{Optimization techniques for low-level \feltor routines}\label{sec:optimization}

As mentioned in Section~\ref{sec:feltor}, the Level 1 algorithms implemented in \feltor include basic linear algebra routines that build the dg library code.
Besides trivially parallel vector operations like addition or pointwise multiplication, we implemented the scalar product with long accumulators (Section~\ref{sec:reproducibility}) and a sparse matrix-vector multiplication.
Optimizing these operations is a key task in order to increase the overall performance of any higher level algorithm or application using \textsc{Feltor}.
%Let us here focus on the optimization of the sparse matrix-vector product for the Knights Landing (KNL) architecture and briefly discuss CPU and GPU implementations thereafter.

Note, that we devised our own sparse block matrix format, which specifically
saves storage on redundant blocks and thus potentially fits into small and fast memory caches of the target architecture. It is used
for the computation of the simple discontinuous Galerkin derivative in $x$ and $y$ on
product spaces (see Reference~\cite{einkemmer2014} for more details).
Many algorithms, including the Arakawa scheme~\cite{einkemmer2014} or the discretization of elliptic problems~\cite{Cockburn1998, Arnold2001}, build on those derivatives. An optimization of the corresponding matrix-vector product will thus greatly contribute to reducing their execution times.

In general, vector additions, sparse-matrix-vector multiplications and scalar products require a similar amount of memory and arithmetic operations. This means that on all modern hardware architectures these routines are memory bound. However, this conclusion assumes an efficient implementation. In particular, it assumes that our code is able to exploit the parallelism present on these architectures in order to saturate the available bandwidth. In addition, to achieve optimal performance, memory has to be read in a sequential (coalesced) manner. This is especially true for the Intel Xeon Phi "Knights Landing" accelerator card (KNL) and GPUs.

The easiest option to optimize a code for a new architecture such as KNL is to recompile it with the proper flags (discussed for KNL further below) and thus get an instantaneous benefit. However, achieving a full and efficient use of a new architecture requires an analysis using available profiling tools and an optimization effort, which is reflected normally in code modifications.

The strategy to optimize a code for a given architecture involves different levels, beginning from the core level to the outer levels of the hardware, since all the optimizations introduced in any level automatically benefits its upper levels.

Most modern processors have so-called \textit{vector units} that allow it to execute a single instruction on multiple data (SIMD) per cycle.
For example, each KNL core  has two 512-bit vector units that enable it to compute 16 double precision operations concurrently.
The usage of these SIMD (or vector) instructions in a loop is called \textit{vectorization}.
%Therefore, vector instructions have to be exploited, otherwise a potential 16x peak speedup could be missed.

Most compilers may vectorize loop structures automatically to take advantage of vector units if they are called with the proper options. For the KNL, the intel compiler  provides \texttt{-xMIC-AVX512} to enable AVX-512 vector instruction set~\cite{avx512}, \texttt{-fma} to generate fused multiply-add (FMA) instructions and \texttt{-align} to use aligned load or store vector instructions.

However, the vectorization report generated by the compiler typically shows that not all loops can be vectorized. The compiler only vectorizes when it considers this process a) safe and b) improves the performance.
This means that in order to achieve a good performance sometimes we have to help the compiler to vectorize loops initially discarded by it.
For example, when the compiler believes that two pointers in a loop may reference a common memory region implying likely data dependencies among iterations the compiler refrains from vectorization.
This situation can be solved using the keyword \texttt{restrict} for a pointer argument in a C/\CC function, which indicates that the pointer argument provides exclusive access to the memory referenced in the function and no other pointer can access it.

Another example is when the compiler does not vectorize a loop because an efficiency heuristics predicts that this vectorization will lead to a worsening of the performance, such as the presence of many unaligned data accesses.
This time it can be solved introducing the OpenMP-4 extension \texttt{\#pragma omp simd}, which explicitly tells the compiler that it is safe to use SIMD instructions. As an example of its application, the following loop in the code reduced the total number of executed instructions by 86\% thanks to enabling SIMD instructions.
\begin{verbatim}
#pragma omp parallel for simd
for (unsigned u = 0; u < size; ++u)
  y_ptr[u] = alpha * x_ptr[u] + beta * y_ptr[u]; 
\end{verbatim}
In general we observe that vectorization significantly improves the performance
of the scalar product with long accumulators,
our sparse matrix-vector multiplication and to a lesser extent also the vector additions.

Continuing with the higher hardware level, a KNL node contains 68 cores, so a good thread scalability is mandatory to take advantage of them.
In the case of the sparse-matrix vector multiplication, the previous code contained three consecutive OpenMP parallel regions that were merged into one to give all threads more work reducing idle time and overhead costs, such as thread management and synchronization.
Besides, KNL offers hyperthreading, which means that each core supports up to $4$ threads, leading to the possibility of using up to $272$ threads per KNL node. As hyperthreading may improve performance when memory access latency limits the execution, some performed experiments suggested to run at least $2$ threads per core in order to increase the full core usage and so improve performance.

Finally, we observe that making the number of polynomial coefficients a compile
time constant (a template parameter) resulted in another significant improvement
of runtime in the matrix-vector multiplication. The coefficient fixes the size of the blocks in the sparse matrix format
and thus the size of the tight inner loops of the routine is known at compile time which allows the compiler to generate more efficient code. This is a common technique that has been utilized in a range of C++ implementations (see, for example, \cite{eigenweb,blazeweb,einkemmer2016high,einkemmer2016resistive}).
%We instantiate the template for typical numbers ($<6$) and else reroute to a default implementation.

Note that all the optimizations that have been performed for the KNL have a positive effect on the regular CPU performance as well. %Although, the effect is significantly smaller in the latter case. (Matthias: this is not true, the vectorization has a similar effect on regular CPUs)
On GPUs we observe similar performance improvements when we add the \texttt{restrict} keyword
to pointer arguments in the corresponding kernels and use template arguments as well.
We can avoid warp divergence since if-clauses are absent in our implementations.

%%%%%%%%%%%%%%%%%%%%%%%%%%%Matthias%%%%%%%%%%%%%%%%%%%%%%%%%%%%%%%%%%%%%%%%%%%%%

\subsection{Configuration}\label{sec:configuration}
We use the program
\verb!feltor/inc/dg/cluster_mpib.cu!
that is contained in~\cite{Feltor-v5.0}, together with suitable submit scripts in~\cite{performance}, for generating the performance data.
Essentially, we gather the average run times of a variety of primitive functions, the Arakawa algorithm and a conjugate gradient iteration.
Let us note here that the results from different architectures are bitwise identical
as long as we only compare results from the same compiler (see Section \ref{sec:reproducibility}).
 We vary problem sizes and number of compute nodes on a selection of representative hardware architectures,
which includes a current consumer grade desktop CPU and GPU, as well as
 dedicated high performance compute hardware from Intel and Nvidia.
Please find a short description of the configuration in
Table~\ref{tab:hardware} and
more details in the dataset~\cite{performance}.
We refer to the documentation of the
\verb!dg::Timer! class in \cite{Feltor-v5.0} for details of how we measure the time on the various architectures involved.

%Also note that
 %we received support from Intel and Nvidia as pointed out in the Acknowledgements.
\begin{table*}[ht]
\begin{centering}
\footnotesize
\input{plots/hardware.tex}
\caption{Description of the tested compute nodes: device description and total RAM size
as well as the corresponding distribution of MPI tasks and OpenMP threads/GPU contexts.
    The configuration of multiple nodes scales the number of MPI tasks; for example $4$ skl nodes involve $8$ MPI tasks each spawning $24$ OpenMP threads running on $192$ physical cores. }
\label{tab:hardware}
\end{centering}
\end{table*}

%In general, we aim to compare the performance using the node/platform/accelerator card fully
%rather than normalizing scalings to the sequential execution time.
%This has several reasons.
%%clusters charge for nodes, single-core version is unoptimized
%On one hand, the current policy of HPC clusters we have access to is to charge
%for \textit{node-usage} rather
%than \textit{core-usage}. A user therefore has an interest in fully using the available resources on a
%given node as efficiently as possible. In fact, we have never run a relevant \feltor
%simulation on just a single CPU core and the single core version of \feltor exists
%mainly for debugging purposes.
%Another reason is that for example on graphics cards there are thousands of threads running in
%parallel and there is no single-threaded execution, nor is there a meaningful way to study
%the scaling of the execution time with the number of threads. With the advent of hyperthreading
%and accelerator cards like the Intel Xeon Phi, this point becomes relevant also for
%hardware other than GPUs.
%We therefore set up MPI tasks, OpenMP threads and GPU streams in a way that optimally uses
%all resources on a given node. We then investigate the scaling to up to 4 nodes of the same
%type and finally compare to other node architectures.

\subsection{Performance measurements}\label{sec:measurements}
From the measured runtime $t$ and the array size $S$
we compute the memory bandwidth $b$ of an algorithm or function
\begin{align}
 %\frac{b}{\text{GB/s}} = 10^{-3}\left(\frac{mS}{\text{MB}}\right)\left(\frac{t}{\text{s}}\right)^{-1}
  b = \frac{mS}{t}
\end{align}
where $m$ is the number of memory loads and stores.
We follow the {STREAM} conventions in counting memory operations, which means that
we separately count each read and each write of a memory location.
For example the vector addition \texttt{axpby}, which computes the operation
$y\leftarrow \alpha x + \beta y$, counts as $m=3$ times the vector size
since we have to read both $x$ and $y$ and then write into $y$. The \texttt{dot}
product $x\cdot y$ counts as $m=2$ times the vector size.

%%%%%%%%%%%%%%%%%%%%%%%%%%%%%%%%%%%%%%%%%%%
\begin{figure}[htpb]
  \centering
\subfloat[axpby]{\includegraphics[width=0.5\textwidth]{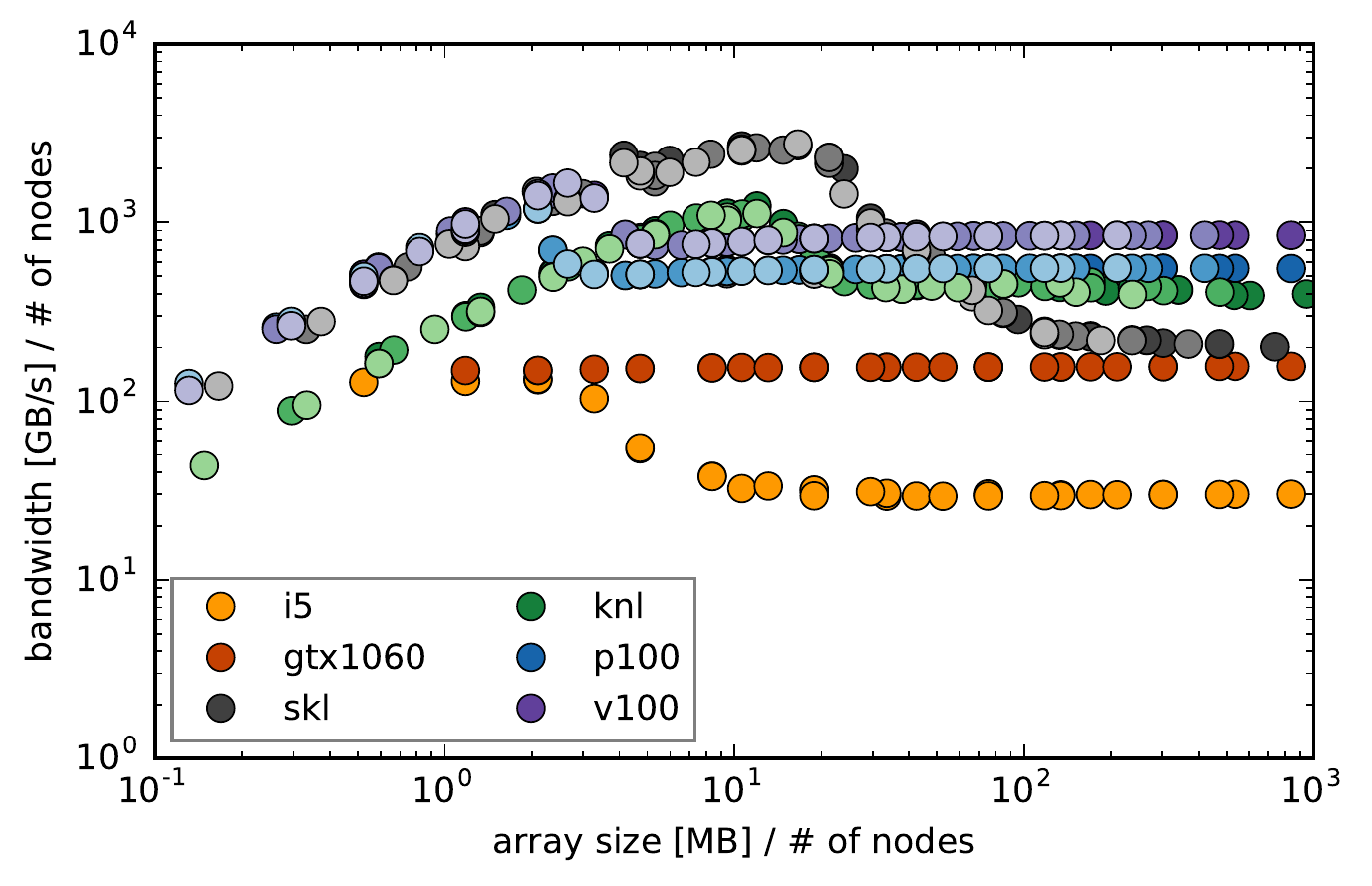} \label{fig:axpby}}  \\
\subfloat[dot]{\includegraphics[width=0.5\textwidth]{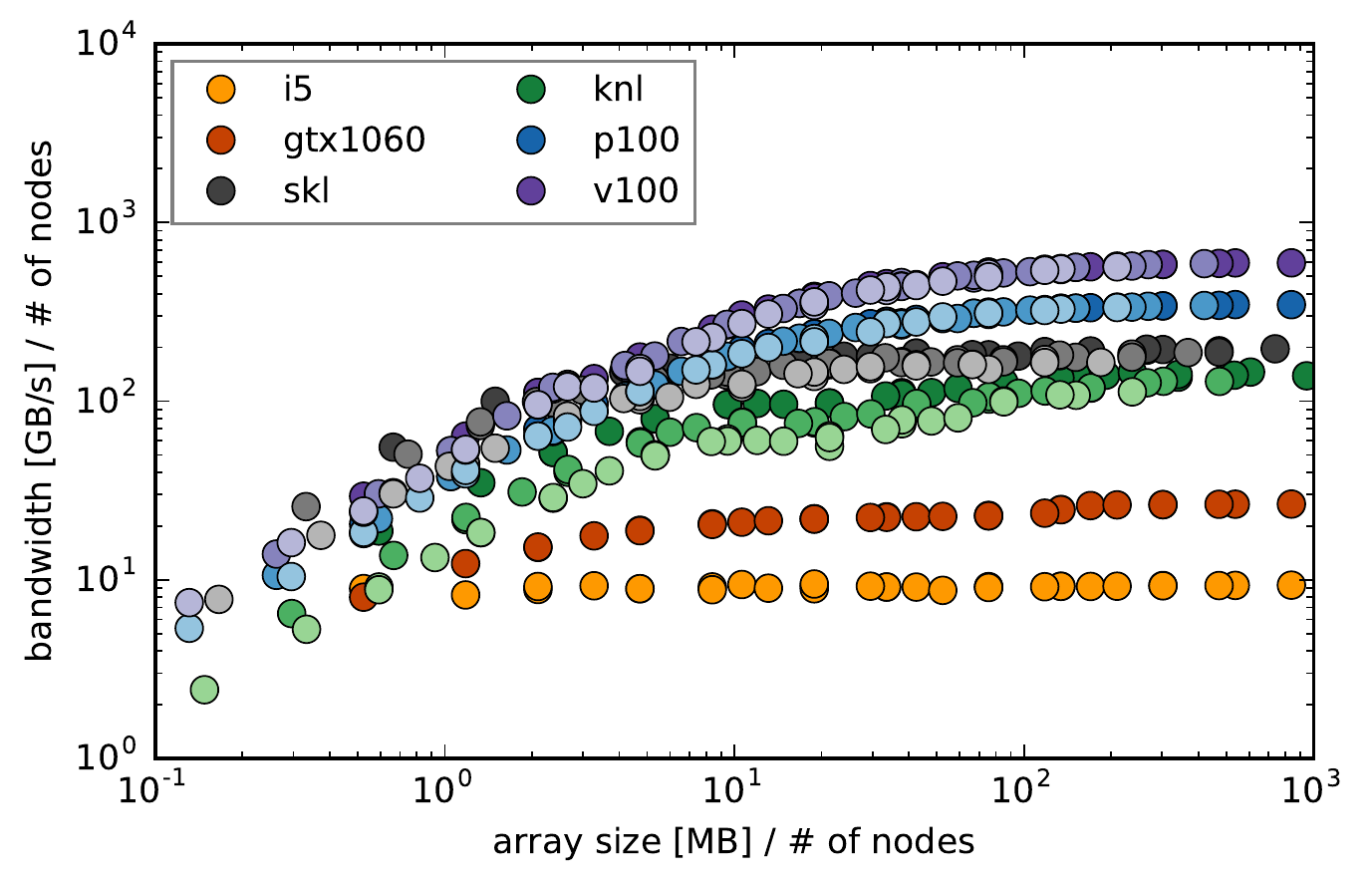} \label{fig:dot}}
  \caption{Single node memory bandwidth plot of the trivially parallel vector addition (a) and
the exact dot product (b) on various hardware architectures on one node (full saturation color),
on two nodes (medium saturation) and on four nodes (low saturation).
We normalize to the number of nodes, such that the single node performance becomes visible.
  }
  \label{fig:dot-performance}
\end{figure}
In Fig.~\ref{fig:dot-performance} we plot the average bandwidth $b$ for various
hardware architectures and problem sizes $S$. We normalize the plot to the number of
nodes $n$, $b/n$ and $S/n$, such that each point represents the performance of a single node.

First, we note that in both, Fig.~\ref{fig:axpby} and Fig.~\ref{fig:dot}
the lightly colored points from multi-node runs lie almost exactly on top
of their single-node counterparts. This is especially true for the P100 and V100 GPUs 
and the Skylake nodes but is not so well fulfilled for the Xeon Phi.
The feature indicates a high \textit{weak scaling efficiency} of both \texttt{axpby} and \texttt{dot}, which means that the achievable bandwidth of a given node is given solely by the problem
size on the node itself. 

Next, we note that in Fig.~\ref{fig:axpby} the
bandwidth for small to medium sized problems ($1$MB $<S/n<10$MB) is significantly higher than for large problems ($S/n>100$MB) for all architectures
(note that the lowest sized point of the 'gtx1060' is hidden beneath a 'skl' point).
This is especially pronounced for the Skylake architecture.
We explain this by the cache level hierarchy. The problem fits
entirely into the cache such that its higher speed becomes visible. In fact, the
peaks roughly coincide with the relevant cache sizes (see \cite{performance} for exact cache sizes).
This feature is absent in Fig.~\ref{fig:dot}, which is most likely due to the high number of 64-bit operations per memory load in the long accumulator scalar product.

For the multi-node architectures  we identify a linear regime for small array sizes ($S/n<2$MB) in
both Fig.~\ref{fig:axpby} and \ref{fig:dot}. Here, the bandwidth
increases linearly with the array size, which indicates a size-independent
runtime $T_\text{lat}$, called the \textit{latency}. Note that the multi-node results
in Fig.~\ref{fig:dot} indicate an increasing latency for multiple nodes.
 We explain this by the necessary global communication between
nodes due to the reduction, which is absent in the \texttt{axpby} algorithm.
On the other hand, for large array sizes ($S/n>100$MB) we identify a regime with constant
bandwidth $B$ in both Fig.~\ref{fig:axpby} and \ref{fig:dot} independent of node number.

\begin{table*}[ht]
\begin{centering}
\footnotesize
\input{plots/axpby-dot.tex}
\caption{Measured single node bandwidth and latency on a single node and four nodes
 of \texttt{apxby} and \texttt{dot}.
 The peak bandwidth is the theoretical RAM bandwidth according to the vendors. The exact peak bandwidth of the MCDRAM on knl was
 not disclosed to us.
}
\label{tab:axpby-dot}
\end{centering}
\end{table*}
We determine $B$ by taking the average bandwidth of the largest array sizes and
estimate an error with the standard deviation.
This is in general a very robust method and yields small errors in our experience.
The correct determination of the latency $T_\text{lat}$ is more involved with the
available data.
 We differentiate between single-node
and multi-node latency.
As a first approximation we simply identify the minimum average runtime with $T_\text{lat}$ and again use the standard deviation as an error estimate.
%Unfortunately, the array sizes that we investigate are too large for the desktop
%hardware 'i5' and 'gtx1060' and the linear regime is not reached.
If we now assume that the runtime is given by
\begin{align}
  t = T_\text{lat}(n) + \frac{mS}{nB},
  \label{eq:primitive_runtime}
\end{align}
then we can correct
the minimum runtime $t_\text{min}$ by $-mS/B$ with the previously measured $B$
to obtain a better approximation to $T_\text{lat}$.
However, with the exception of the Knights landing
architecture the single-node latencies for \texttt{axpby} are so small that the
values become negative. In this case we replace the value by $0$.
In Table~\ref{tab:axpby-dot} we give numerical values of the bandwidths together with the latencies as well as the peak bandwidth according to the vendors.
Within the error the \texttt{axpby} latencies can be neglected altogether
 except for the
Knights Landing architecture.

We note that the GPUs and the Xeon Phi have the highest latencies in the \texttt{dot}
algorithm. The high GPU latency is the result of the slow PCIe lanes since the
result has to be sent back to the host CPU, which entails communication.
As already evident in Figure~\ref{fig:dot} the latencies on multiple nodes significantly increase for the Xeon Phi and the Skylake architectures. This indicates a
long latency of the internode connection. For the P100 and V100 GPUs the
latency seems to be dominated by the communication between GPU and host CPU.

Finally, we note that all architectures reach only $75\%$ (p100) to $95\%$ (v100) of their theoretical peak bandwidth in the \texttt{axpby} function.
This is in line with previous observations of the STREAM benchmark~\cite{Deakin2016}.
We do not know of any practical method to overcome this performance degradation programmatically
and consider the measured bandwidth $B^{\texttt{axpby}}$ as the maximum bandwidth any memory bound algorithm can achieve on the
given architecture. In this sense, the Skylake architecture reaches almost $100\%$
efficiency for the \texttt{dot} algorithm, followed by the Tesla cards P100 and V100.
The GTX 1060 has the lowest efficiency, which is most likely due to the drastic reduction of
double precision performance on the gaming GPU (a factor $32$ compared to single precision), which is absent in the Tesla GPUs.

%\begin{figure}[htpb]
%  \centering
%\subfloat[CPUs]{\includegraphics[width=0.9\textwidth]{./plots/dxdy_cpu.pdf}\label{fig:dxdy_cpu}} \\
%\subfloat[GPUs]{\includegraphics[width=0.9\textwidth]{./plots/dxdy_gpu.pdf}\label{fig:dxdy_gpu}}
%  \caption{Comparison of the matrix-vector multiplication for three 
%polynomial coefficients and various array sizes for host CPUs (a) and GPUs (b). 
%We compare the normalized results on a single node (high saturation color) to 2 (medium saturation color) and 4 nodes (low saturation color).
%The latency dominated regimes for MPI are clearly visible for small array sizes. }
%  \label{fig:dxdy}
%\end{figure}
\begin{table*}[ht]
\begin{centering}
\footnotesize
\input{./plots/dxdy.tex}
\caption{Single node bandwidth of a DG matrix-vector multiplication for various
polynomial coefficients $P$. Latencies on a single node/card and four nodes/cards.
}
\label{tab:dxdy}
\end{centering}
\end{table*}
For the matrix-vector product we perform the same analysis and show
the results In Table~\ref{tab:dxdy}. There, we present the average
bandwidth between a DG derivative in the x-direction and the y-direction, which
we call \texttt{dxdy}.
Due to our efficient format the matrix itself does not contribute to the memory loads and stores. We count two
loads and one store for the $y\leftarrow\alpha M x+\beta y$ operation ($m=3$).
However, we do differentiate between various polynomial orders, which determines
the \textit{stencil} of the operation. A higher polynomial order increases the
registry pressure and thus decrease the efficiency of the implementation. 
The latency should not be influenced by the polynomial order and
we provide the latencies for the $P=2$ case.
We observe the highest latencies for the multi-node configurations. Here, the
algorithm involves communication between neighboring processes. This is particularly
unfavourable for the GPUs since these have to communicate via the host CPU
across the PCIe lanes.

Concerning the single node bandwidths $B$ we overall observe the highest values for the Tesla GPUs.
It is noteworthy that the GTX 1060 has a very low latency and reaches almost $70\%$ of the bandwidth of the
much more expensive Skylake and Xeon Phi nodes.
%%%%%%%%%%%%%%%%%%%%%%%%%%%%%%%%%%%%%%%%%%%%%%%%%%%%%%%%%%%%%%
\subsection{Performance prediction model}\label{sec:prediction}
Any one of the primitive subroutines in Level $1$ in \feltor falls into
one of the categories 'trivially parallel' (\texttt{axpby}), 'nearest neighbor communication' (\texttt{dxdy}) and 'global communication' (\texttt{dot}).
We specifically measured the bandwidths and latencies
of the three operations \texttt{axpby}, \texttt{dxdy} and \texttt{dot} in Tables~\ref{tab:axpby-dot} and \ref{tab:dxdy}.
For the following discussion we assume that these values
accurately represent the bandwidths and latencies of the whole respective class of functions.
In fact, we use these values to predict the runtime of any algorithm that is
implemented in terms of Level $1$ subroutines.
For a given architecture and node number $n$
we predict a runtime $t$ depending on the array size $S$ and the
number of polynomial coefficients $P$
\begin{align}
  t(P,S,n) &= \sum_q \sum_{i=0}^{N_q-1} t_i^q(P,S,n) = \sum_q \left[N_q T^q_\text{lat}(n) + \frac{M^q S}{nB^q(P)}\right] \nonumber \\
  &=: N\left[ T_\text{lat}(n) + \frac{M}{N}\frac{ S}{n B (P)}\right] \label{eq:runtime}\\
T_\text{lat}(n) &:= \frac{1}{N}\sum_q N_q T^q_\text{lat}(n)\label{eq:avg_latency}\\
\frac{1}{B(P)} &:= \frac{1}{M}\sum_q \frac{M^q}{B^q(P)}  \label{eq:avg_bandwidth}
\end{align}
with the function type $q\in(\texttt{axpby,dot,dxdy})$, $i$ iterates over all
occurrences of function type $q$,
$N_q$ is the total number of occurrences of all functions of
type $q$, $M^q$ is the total
number of memory loads and stores among functions of type $q$,
$B^q(P)$ is the single node memory bandwidth of function type $q$
and $T^q_\text{lat}(n)$ is the latency depending on the
number of nodes used. In Eqs.~\eqref{eq:avg_latency} and \eqref{eq:avg_bandwidth} we defined the average latency and weighted average single node bandwidth,
where $N:=\sum_qN_q$ and $M:=\sum_qM^q$.
The values for $B^q(P)$ and $T^q_\text{lat}(n)$ are in Table~\ref{tab:axpby-dot} and \ref{tab:dxdy}. We present an average over a conjugate gradient iteration and 
the Arakawa algorithm in Table~\ref{tab:avg}. These two algorithms represent a 
typical mixture of primitive functions used in a \feltor simulation project.
In fact, we get a first approximation of the runtime of any algorithm 
by counting the total 
number of function calls $N$ and using Eq.~\eqref{eq:runtime}, Table~\ref{tab:avg}
and $M/N\approx 3.3$.

\begin{table*}[ht]
\begin{centering}
\footnotesize
\input{plots/avg.tex}
\caption{
Average single node bandwidths $B$ for various polynomial coefficients $P$ as well as average single-node
  and multi-node latencies according to Eq.~\eqref{eq:runtime}.
  We use Table~\ref{tab:axpby-dot} and \ref{tab:dxdy}, $(N^\texttt{axpby},N^\texttt{dot},N^\texttt{dxdy})=(9,2,12)$ and $(M^\texttt{axpby},M^\texttt{dot},M^\texttt{dxdy})=(36,4,36)$, $N=23$, $M=76$ and a ratio of $M/N=3.30$.
  This corresponds to the average between a conjugate gradient iteration and the Arakawa algorithm.
}
\label{tab:avg}
\end{centering}
\end{table*}
%%%%%%%%%%%%%%%%%%%%%%%%%%%%%%%%%%%%%%%%%%%%%%
\begin{figure}[htpb]
  \centering
\subfloat[arakawa]{\includegraphics[width=0.5\textwidth]{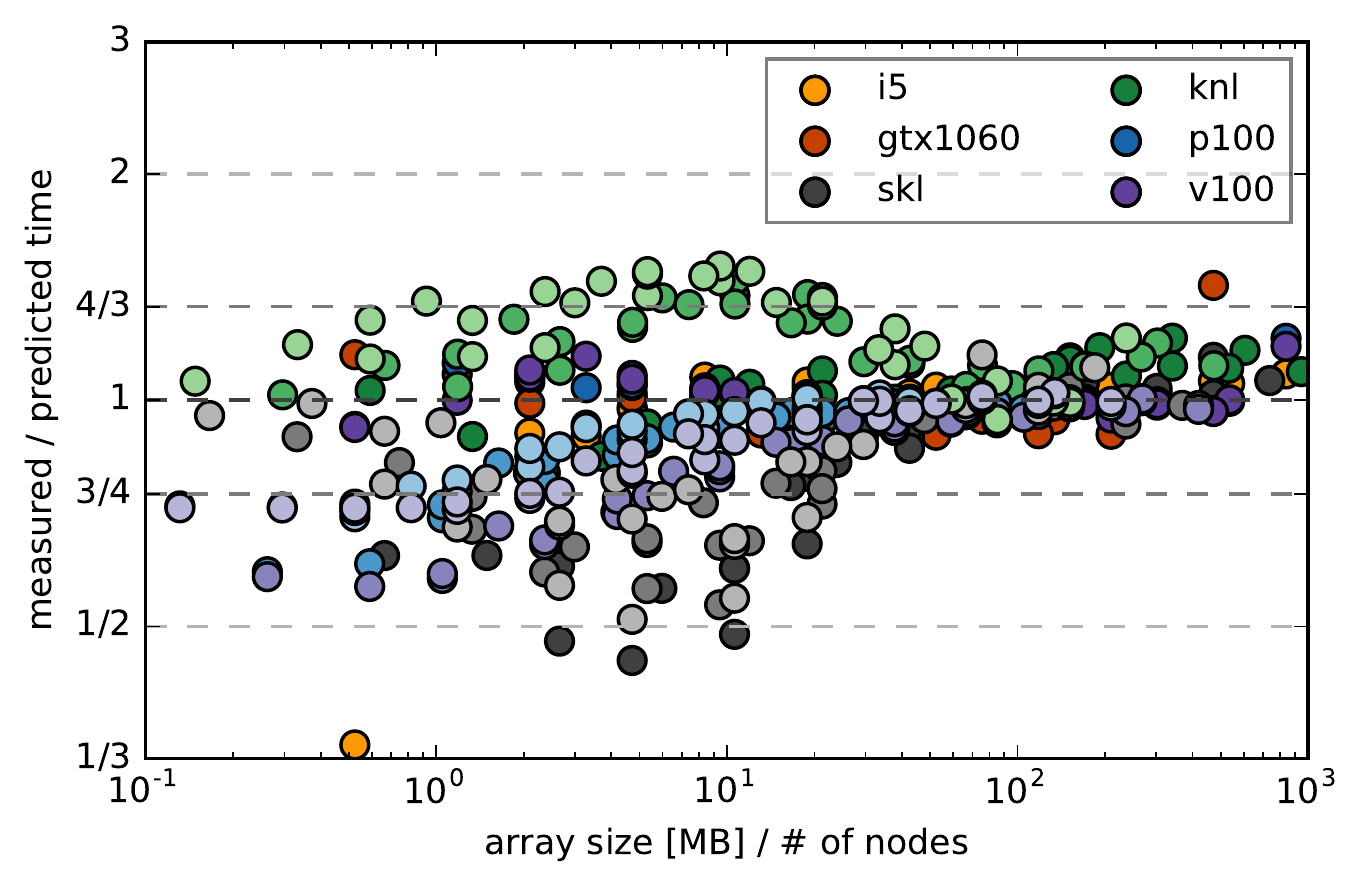}\label{fig:arakawa}} \\
\subfloat[cg]{\includegraphics[width=0.5\textwidth]{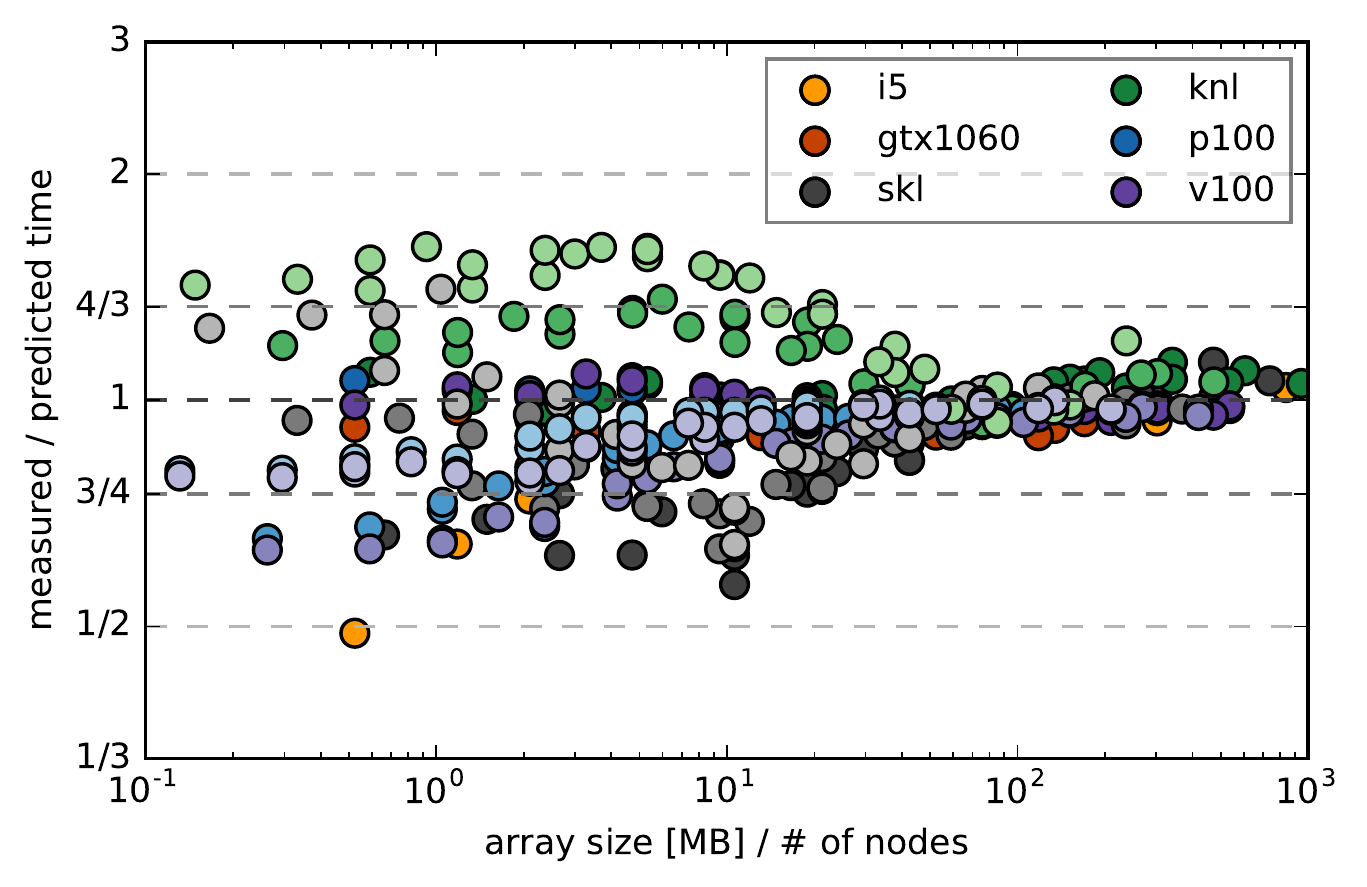}\label{fig:cg}}
\caption{Comparison of predicted runtime Eq.~\eqref{eq:runtime} to measured time for the Arakawa
algorithm (a) and a single conjugate gradient (cg) iteration (b). The
plot highlights the deviation from the predicted time. Points below $1$ mean
faster execution than predicted.
}
  \label{fig:prediction}
\end{figure}
In Fig.~\ref{fig:prediction} we compare the result of the prediction in Eq.~\eqref{eq:runtime} with the measured runtime for the \texttt{arakawa} algorithm and one \texttt{cg} iteration.
The plots depict the relative error of the prediction. If a point lies below $1$,
the execution was faster than predicted.
Especially for large array sizes $S/n>30$MB our prediction is accurate for all
architectures.
We note in both Fig.~\ref{fig:arakawa} and Fig.~\ref{fig:cg} that the measured runtime for the Xeon Phi card
on multiple nodes for sizes $S/n<20$MB is systematically overestimated.
On the other side the Skylake architecture and the Intel i5 CPU for array
sizes $S/n<30$MB run up to a factor $2$ faster than predicted.
We explain this by the very fast execution of the trivially parallel
part of the algorithms in the fast cache as is evident in Fig.~\ref{fig:axpby}.
This effect is not included in our parallel model.
On the other hand, the measured run times $T^\text{meas}$ for the Tesla GPUs and our desktop system are remarkably well predicted by our model and with only few exceptions lie within an interval $(3/4)T^\text{pred} < T^\text{meas} < (4/3)T^\text{pred}$, with $T^\text{pred}$ given
by Eq.~\ref{eq:runtime}.

\subsection{Strong and weak scaling}\label{sec:scaling}
Equation~\eqref{eq:runtime} enables us to discuss the strong and weak scaling
of an arbitrary algorithm. The strong scaling of a problem with total array size $S$,
polynomial coefficients $P$ and number of nodes $n$ is defined as
\begin{align}
  \varepsilon(P,S,n) := \frac{t(P,S,1)}{nt(P,S,n)}=\frac{T_\text{lat}(1) + (M/N)(S/B(P))}{nT_\text{lat}(n) + (M/N)(S/B(P))}
  \label{}
\end{align}
The weak scaling efficiency relates run times with equal array size per node $s=S/n$ as
\begin{align}
  \gamma(P,s,n) := \frac{t(P,s,1)}{t(P,ns,n)} = \frac{T_\text{lat}(1) + (M/N)(s/B(P))}{T_\text{lat}(n) + (M/N)(s/B(P))}
  \label{}
\end{align}
We immediately see that the efficiency $\varepsilon$ tends to $0$ for large number of nodes $n$, while the explicit $n$ dependency in $\gamma$ vanishes.
The only remaining dependence on $n$ is in the latency $T_\text{lat}(n)$.
We argue that the dependence on $n$
should vanish in the latencies for \texttt{axpby}, since there is no
communication at all. For the \texttt{dxdy} algorithm the latencies should also
become independent of $n$ for large $n$ since communication happens only
between nearest neighbors. Only for the \texttt{dot} product the latency should
increase with $n$
%$\sqrt{n}$for large $n$
due to the global communication.

Both the strong and the weak scaling tend to unity if $s=S/n\gg (N/M)T_\text{lat}(n)B(P)$. The value of the product
\begin{align}
  (S/n)_\text{min} \approx 0.3 T_\text{lat}(n)B(P)
  \label{eq:minimum_size}
\end{align}
is the minimum size per node for a \feltor simulation
with a scaling efficiency of at least $50\%$ (both $\varepsilon$ and $\gamma$
are $>0.5$ for $S/n > (S/n)_\text{min}$).
For the values presented in Table~\ref{tab:avg} the minimum array size per node
typically lies between $1$ and $10$MB.

Note that for fixed latencies the scaling efficiencies $\varepsilon$
and $\gamma$ are low if the bandwidth is high and
vice versa high if the bandwidth is low.
This is ironical since the runtime $t$ is low if the bandwidth is high as
evident in Eq.~\eqref{eq:runtime}.
A high scaling efficiency is thus not necessarily an indicator
for an efficient implementation just as a low scaling efficiency not necessarily
points to inefficient code.
%We conclude that the strong/weak scaling efficiency
%fails to correctly capture the efficiency of a (memory bandwidth bound)
%implementation.

\subsection{Discussion}\label{sec:performance_discussion}
From Eq.~\eqref{eq:runtime} it is clear that the runtime $t$ is low if
the average latency $T_\text{lat}$ is low and the bandwidth $B$ is high.
On the other side, performance can also be gained by reducing the number of function calls $N$, or the number of memory operations $M$.
Apparently, the fastest possible implementation is to implement the whole algorithm
in a single function, with $N=1$ and a minimum number of
memory operations $M_\text{min}$. For example, our current \texttt{arakawa} implementation
has $M=34$ and $N=9$, which compares unfavourably to a possible $N_\text{min}=1$ and $M_\text{min}=4$.
The drawback of implementing and optimizing every algorithm or equation
separately is the increased maintenance and performance tuning cost.
Furthermore, this approach would not be easily extensible or modifiable
and violates our design goals presented in Section~\ref{sec:feltor}.
Still, we estimate the performance we loose due to the \feltor design between
a factor $2$ and $5$ depending on the algorithm at hand.
In an effort to mitigate the problem we introduced new primitive
functions with increased workload,
for example the vector operation $z\leftarrow \alpha x_1y_1+\beta x_2y_2 + \gamma z$,
where $x_1y_1$ is a point-wise multiplication.
%Since the performance is determined by the number of memory loads and stores (bandwidth) and
%the number of function calls (latency)
%we want to further explore possibilities to reduce those in the future.
We currently also explore template parameter packs in the \CC-11 standard as
a promising candidate to increase the workload of Level 1 functions in \textsc{Feltor}.

%% file: plots/hardware.tex
\begin{tabular}{@{}lp{6.5cm}p{5cm}@{}}
\toprule
{} &                                                         device description &                                    single-node configuration \\
\midrule
\textbf{i5     } &                        Intel Core i5-6600 @ 3.30GHz (2x 8GB DDR4, 4 cores) &                   1 MPI task x 4 OpenMP threads (1 per core) \\
\textbf{skl    } &      2x Intel Xeon 8160 (Skylake) at 2.10 GHz (12x 16GB DDR4, 2x 24 cores) &  2 MPI tasks (1 per socket) x 24 OpenMP threads (1 per core) \\
\textbf{knl    } &  Intel Xeon Phi 7250 (Knights Landing) at 1.40 GHz (16GB MCDRAM, 68 cores) &            1 MPI task x 136 OpenMP hyperthreads (2 per core) \\
\textbf{gtx1060} &                                Nvidia GeForce GTX 1060 (6GB global memory) &                                           1 MPI task per GPU \\
\textbf{p100   } &                                Nvidia Tesla P100-PCIe (16GB global memory) &                                           1 MPI task per GPU \\
\textbf{v100   } &                                Nvidia Tesla V100-PCIe (16GB global memory) &                                           1 MPI task per GPU \\
\bottomrule
\end{tabular}

%% file: plots/axpby-dot.tex
\begin{tabular}{@{}lp{1.5cm}p{1.5cm}p{1.2cm}p{1.2cm}p{1.5cm}p{1.2cm}p{1.4cm}@{}}
\toprule
{} &             peak & \multicolumn{3}{l}{axpby} & \multicolumn{3}{l}{dot} \\
{} & bandwidth [GB/s] & bandwidth [GB/s] & $T_{lat}(1)$ [$\mu$s] & $T_{lat}(4)$ [$\mu$s] & bandwidth [GB/s] & $T_{lat}(1)$ [$\mu$s] & $T_{lat}(4)$ [$\mu$s] \\
\midrule
\textbf{i5     } &               34 &      30 $\pm$ 01 &           00 $\pm$ 02 &                   n/a &      10 $\pm$ 01 &           05 $\pm$ 01 &                   n/a \\
\textbf{gtx1060} &              192 &     158 $\pm$ 01 &           00 $\pm$ 01 &                   n/a &      27 $\pm$ 01 &           93 $\pm$ 09 &                   n/a \\
\textbf{skl    } &              256 &     207 $\pm$ 06 &           00 $\pm$ 01 &           00 $\pm$ 01 &     193 $\pm$ 19 &           18 $\pm$ 03 &           38 $\pm$ 05 \\
\textbf{knl    } &             >400 &     394 $\pm$ 23 &           06 $\pm$ 01 &           10 $\pm$ 01 &     142 $\pm$ 07 &           55 $\pm$ 02 &          120 $\pm$ 06 \\
\textbf{p100   } &              732 &     554 $\pm$ 01 &           01 $\pm$ 01 &           03 $\pm$ 01 &     347 $\pm$ 02 &           49 $\pm$ 01 &           49 $\pm$ 01 \\
\textbf{v100   } &              898 &     849 $\pm$ 01 &           02 $\pm$ 01 &           03 $\pm$ 01 &     594 $\pm$ 03 &           34 $\pm$ 02 &           35 $\pm$ 01 \\
\bottomrule
\end{tabular}

%% file: plots/dxdy.tex
\begin{tabular}{lp{1.5cm}p{1.5cm}p{1.5cm}p{1.5cm}p{1.2cm}p{1.2cm}}
\toprule
{} & B(P=2) [GB/s] & B(P=3) [GB/s] & B(P=4) [GB/s] & B(P=5) [GB/s] & $T_{lat}(1)$ [$\mu$s] & $T_{lat}(4)$ [$\mu$s] \\
\midrule
\textbf{i5     } &   28 $\pm$ 03 &   30 $\pm$ 03 &   26 $\pm$ 02 &   22 $\pm$ 02 &           00 $\pm$ 02 &                   n/a \\
\textbf{gtx1060} &  131 $\pm$ 01 &  112 $\pm$ 02 &   84 $\pm$ 14 &   70 $\pm$ 18 &           00 $\pm$ 01 &                   n/a \\
\textbf{skl    } &  182 $\pm$ 36 &  162 $\pm$ 13 &  119 $\pm$ 19 &  111 $\pm$ 09 &           23 $\pm$ 03 &           29 $\pm$ 03 \\
\textbf{knl    } &  240 $\pm$ 18 &  173 $\pm$ 27 &  127 $\pm$ 19 &  102 $\pm$ 15 &           10 $\pm$ 01 &           53 $\pm$ 04 \\
\textbf{p100   } &  288 $\pm$ 03 &  238 $\pm$ 04 &  201 $\pm$ 02 &  166 $\pm$ 15 &           02 $\pm$ 01 &           64 $\pm$ 01 \\
\textbf{v100   } &  802 $\pm$ 17 &  713 $\pm$ 20 &  650 $\pm$ 16 &  536 $\pm$ 49 &           04 $\pm$ 01 &           67 $\pm$ 02 \\
\bottomrule
\end{tabular}

%% file: plots/avg.tex
\begin{tabular}{lp{1.5cm}p{1.5cm}p{1.5cm}p{1.5cm}p{1.2cm}p{1.2cm}}
\toprule
{} & B(P=2) [GB/s] & B(P=3) [GB/s] & B(P=4) [GB/s] & B(P=5) [GB/s] & $T_{lat}(1)$ [$\mu$s] & $T_{lat}(4)$ [$\mu$s] \\
\midrule
\textbf{i5     } &   26 $\pm$ 02 &   27 $\pm$ 02 &   26 $\pm$ 01 &   23 $\pm$ 02 &           01 $\pm$ 01 &                   n/a \\
\textbf{gtx1060} &  116 $\pm$ 01 &  108 $\pm$ 01 &   94 $\pm$ 09 &   85 $\pm$ 12 &           09 $\pm$ 01 &                   n/a \\
\textbf{skl    } &  194 $\pm$ 20 &  183 $\pm$ 09 &  153 $\pm$ 15 &  147 $\pm$ 07 &           14 $\pm$ 02 &           19 $\pm$ 02 \\
\textbf{knl    } &  281 $\pm$ 13 &  232 $\pm$ 24 &  188 $\pm$ 20 &  160 $\pm$ 18 &           13 $\pm$ 01 &           42 $\pm$ 02 \\
\textbf{p100   } &  377 $\pm$ 02 &  333 $\pm$ 04 &  297 $\pm$ 02 &  259 $\pm$ 17 &           06 $\pm$ 01 &           39 $\pm$ 01 \\
\textbf{v100   } &  808 $\pm$ 09 &  763 $\pm$ 11 &  727 $\pm$ 10 &  653 $\pm$ 35 &           06 $\pm$ 01 &           39 $\pm$ 01 \\
\bottomrule
\end{tabular}

%% file: conclusion.tex
\section{Discussion and Conclusion}\label{sec:conclusion}
%In this contribution we explicitly show the interconnection between physical model equations,
%  numerical analysis, implementation,
%  performance optimizations and computational hardware
%  in our numerical simulations.
%  We achieved bitwise reproducibility. Then we discussed that in order to judge the
%  quality of a simulation only reduced physical quantities of interest or invariants
%  may show convergence.
%  This promotes the use of physical models that show invariants and numerical
%  methods that conserve them. One example of such a method is Arakawa's algorithm
%  to discretize the Poisson bracket. We compare two high order variants of the algorithm and find that the discontinuous Galerkin approach shows less numerical diffusion
%  and reproduces the fine structure more faithfully than the original finite
%  difference method.
With Table~\ref{tab:avg} and Eq.~\ref{eq:runtime} in Section~\ref{sec:performance} we have a powerful tool to judge the performance
of various hardware architectures available to us. We are
able to estimate the runtime of a \feltor simulation for given problem size,
hardware and node count.
Furthermore, with Eq.~\eqref{eq:minimum_size} we have an easy
to use estimate of the minimum array size per compute node for an
acceptable scaling efficiency.
From a user perspective this estimate and the possibility to predict runtimes are certainly valuable
features since available resources can be used more effectively and the performance
of future hardware can be estimated from its theoretical bandwidth.

Let us here discuss performance also in the light of the
simulations we eventually want to carry out.
In fact, an increase/decrease in performance of the implementation
may lead to an only marginal improvement/deterioration of the numerical simulations itself.
Consider for example a three dimensional problem and an available runtime $T$
(set by cluster policies, allocated resources or simply personal preferences).
Accounting for the reduced/increased time step
due to the CFL condition
a factor $2$ increase/decrease in performance leads to only a factor
$2^{1/4}\approx 1.19$ increase/decrease in
the number of grid points per dimension.
This justifies our design goals laid out in Section~\ref{sec:feltor}. We
do strife for performance but when faced with possible trade-off scenarios
we put equal value on other goals as well.

Of course, the choice of the physical model and the numerical methods employed
ultimately set the limit of what an implementation can achieve in terms of performance.
In discontinuous Galerkin methods the order of the method is a free parameter.
In Section~\ref{sec:performance} we argue that a higher order method
executes slower than a lower order method with the same number of degrees of freedom due to the increased stencil.
At the same time,
%in Section~\ref{sec:arakawa} we elaborate that
the high order method may also require less
points overall to achieve the same resolution as the lower order method.
The minimum requirements of what a simulation has to resolve is eventually given by the spatial and temporal scales in the physical dynamics.

Another consideration is the question of when a simulation is ``converged''.
As we argue in Section~\ref{sec:reproducibility} this question can be difficult
to answer.
Algorithmically and programmatically we do achieve accurate and
bitwise reproducible results. We do so by
ensuring deterministic execution of elementary subroutines like the dot product,
which serve as building blocks for our parallel algorithms.
%Section~\ref{sec:performance} shows that our implementation of these routines
%leads to highly performant simulations on various architectures.
However, in ill-conditioned problems
only reduced physical quantities of interest, ensemble simulations or invariants
may be able to indicate correct simulation results.
Pointwise convergence is possible only for reduced simulation times.

We hope that this discussion provides the reader with the tools necessary to justify an
appropriate setup for a numerical simulation
and although this article was primarily written with \feltor in mind we think that
our arguments hold for any similar simulation framework as well.

 %In summary we argue that
 %the physical model equations,
 %the numerical analysis,
 %the implementation,
 %and the computational hardware
 %all have to be considered equally when performing numerical simulations.
%In fact, we argue that the discussion of
%\textit{Exascale} is mostly irrelevant to us since the performance of our
%algorithms is not determined by the floating point operation performance of
%an architecture but rather the memory bandwidth of its RAM.

\section*{Acknowledgements}

We thank Harald Servat, from the Intel Corporation, for the help provided in optimizing \feltor on the Intel Xeon Phi "Knights landing" and Skylake
architectures.
We thank Siegfried H\"ofinger from the Technical University of Vienna
for mediating the contact to NVIDIA Corporation.
We gratefully acknowledge the support of NVIDIA PSG Cluster with the donation of the Tesla P100 and V100 GPUs used for this research.

The research leading to these results has received funding from the European Union's Horizon 2020 research and innovation programme under the Marie Sklodowska-Curie grant agreement no. 713683 (COFUNDfellowsDTU).
This work has been carried out within the framework of the EUROfusion Consortium and has received funding from the Euratom research and training programme 2014-2018 under grant agreement No 633053. The views and opinions expressed herein do not necessarily reflect those of the European Commission.

%% file: paper.bbl
\begin{thebibliography}{10}

\bibitem{acm-repro-initiative}
{ACM Artifact Review and Badging}.
\newblock
  \url{https://www.acm.org/publications/policies/artifact-review-badging}.

\bibitem{acm-toms-rcr}
{ACM TOMS Replicated Computational Results}.
\newblock \url{https://toms.acm.org/replicated-computational-results.cfm}.

\bibitem{Homepage}
Feltor homepage.
\newblock \url{https://feltor-dev.github.io}.

\bibitem{sc-repro-initiative}
{SC Reproducibility Initiative}.
\newblock
  \url{https://sc18.supercomputing.org/submit/sc-reproducibility-initiative/}.

\bibitem{alexandrescu2001}
A.~Alexandrescu.
\newblock {\em {Modern C++ Design}}.
\newblock Addison-Wesley, 2001.

\bibitem{Arnold2001}
D.~N. Arnold, F.~Brezzi, B.~Cockburn, and L.~D. Marini.
\newblock Unified analysis of discontinuous galerkin methods for elliptic
  problems.
\newblock {\em SIAM J. Numer. Anal.}, 39(5):1749--1779, January 2002.

\bibitem{ascac}
S.~Ashby~et al.
\newblock {The opportunities and challenges of exascale computing}.
\newblock {\em Report of the ASCAC Subcommittee on Exascale Computing}, 2010.

\bibitem{baboulin2009}
M.~Baboulin, A.~Buttari, J.~Dongarra, J.~Kurzak, J.~Langou, J.~Langou,
  P.~Luszczek, and S.~Tomov.
\newblock {Accelerating scientific computations with mixed precision
  algorithms}.
\newblock {\em Comput. Phys. Commun.}, 180(12):2526--2533, 2009.

\bibitem{Beer1996}
M.~A. Beer and G.~W. Hammett.
\newblock {Toroidal gyrofluid equations for simulations of tokamak turbulence}.
\newblock {\em Phys. Plasmas}, 3(11):4046, 1996.

\bibitem{Braginskii}
Braginskii.
\newblock {Transport processes in a plasma}.
\newblock {\em Rev. Plasma Phys.}, Vol. 1, 1965.

\bibitem{chen2012}
G.~Chen, L.~Chac{\'o}n, and D.~C. Barnes.
\newblock {An efficient mixed-precision, hybrid {CPU}--{GPU} implementation of
  a nonlinearly implicit one-dimensional particle-in-cell algorithm}.
\newblock {\em J. Comput. Phys.}, 231(16):5374--5388, 2012.

\bibitem{Cockburn1998}
B.~Cockburn and C.~W. Shu.
\newblock {The local discontinuous Galerkin method for time-dependent
  convection-diffusion systems}.
\newblock {\em SIAM J. Numer. Anal.}, 35(6):2440--2463, November 1998.

\bibitem{Collange15Parco}
S.~Collange, D.~Defour, S.~Graillat, and R.~Iakymchuk.
\newblock Numerical reproducibility for the parallel reduction on multi- and
  many-core architectures.
\newblock {\em {Parallel Computing}}, 49:83--97, 2015.

\bibitem{cusp}
S.~Dalton, N.~Bell, L.~Olson, and M.~Garland.
\newblock Cusp: Generic parallel algorithms for sparse matrix and graph
  computations, 2014.
\newblock Version 0.5.0.

\bibitem{Deakin2016}
T.~Deakin, J.~Price, M.~Martineau, and S.~McIntosh-Smith.
\newblock {GPU}-stream v2.0: Benchmarking the achievable memory bandwidth of
  many-core processors across diverse parallel programming models.
\newblock {\em High Performance Computing, Isc High Performance 2016
  International Workshops}, 9945:489--507, 2016.

\bibitem{Dudson2015}
B.~D. Dudson, A.~Allen, G.~Breyiannis, E.~Brugger, J.~Buchanan, L.~Easy,
  S.~Farley, I.~Joseph, M.~Kim, A.~D. McGann, J.~T. Omotani, M.~V. Umansky,
  N.~R. Walkden, T.~Xia, and X.~Q. Xu.
\newblock Bout plus plus : Recent and current developments.
\newblock {\em J. Plasma Phys.}, 81:365810104, January 2015.

\bibitem{einkemmer2016mixed}
L.~Einkemmer.
\newblock {A mixed precision semi-Lagrangian algorithm and its performance on
  accelerators}.
\newblock In {\em High Performance Computing \& Simulation (HPCS), 2016
  International Conference on}, pages 74--80, 2016.

\bibitem{einkemmer2016resistive}
L.~Einkemmer.
\newblock {A resistive magnetohydrodynamics solver using modern C++ and the
  Boost library}.
\newblock {\em Comput. Phys. Commun.}, 206:69--77, 2016.

\bibitem{einkemmer2016high}
L.~Einkemmer.
\newblock {High performance computing aspects of a dimension independent
  semi-Lagrangian discontinuous Galerkin code}.
\newblock {\em Comput. Phys. Commun.}, 202:326--336, 2016.

\bibitem{einkemmer2018perfcomp}
L.~Einkemmer.
\newblock {A performance comparison of semi-Lagrangian discontinuous Galerkin
  and spline based Vlasov solvers in four dimensions}.
\newblock {\em J. Comput. Phys.}, 2018.

\bibitem{einkemmer2014}
L.~Einkemmer and M.~Wiesenberger.
\newblock {A conservative discontinuous Galerkin scheme for the {2D}
  incompressible Navier--Stokes equations}.
\newblock {\em Comput. Phys. Commun.}, 185(11):2865--2873, 2014.

\bibitem{Fasoli2016}
A.~Fasoli, S.~Brunner, W.~A. Cooper, J.~P. Graves, P.~Ricci, O.~Sauter, and
  L.~Villard.
\newblock Computational challenges in magnetic-confinement fusion physics.
\newblock {\em Nature Physics}, 12(5):411--423, May 2016.

\bibitem{Goldberg1991}
D.~Goldberg.
\newblock What every computer scientist should know about floating-point
  arithmetic.
\newblock {\em Computing Surveys}, 23(1):5--48, March 1991.

\bibitem{eigenweb}
G.~Guennebaud, B.~Jacob, et~al.
\newblock {Eigen v3}.
\newblock \url{http://eigen.tuxfamily.org}, 2018.

\bibitem{hairer2006geometric}
E.~Hairer, C.~Lubich, and G.~Wanner.
\newblock {\em {Geometric numerical integration: structure-preserving
  algorithms for ordinary differential equations}}, volume~31.
\newblock Springer, 2006.

\bibitem{Halpern2016}
F.~D. Halpern, P.~Ricci, S.~Jolliet, J.~Loizu, J.~Morales, A.~Mosetto,
  F.~Musil, F.~Riva, T.~M. Tran, and C.~Wersal.
\newblock The gbs code for tokamak scrape-off layer simulations.
\newblock {\em J. Comput. Phys.}, 315:388--408, June 2016.

\bibitem{Hariri2013}
F~Hariri and M~Ottaviani.
\newblock A flux-coordinate independent field-aligned approach to plasma
  turbulence simulations.
\newblock {\em Comput. Phys. Commun.}, 184(11):2419--2429, 2013.

\bibitem{hasegawa83}
A.~Hasegawa and M.~Wakatani.
\newblock {Plasma edge turbulence}.
\newblock {\em Phys. Rev. Lett.}, 50:682, 1983.

\bibitem{wakatani87}
A.~Hasegawa and M.~Wakatani.
\newblock Self-organization of electrostatic turbulence in a cylindrical
  plasma.
\newblock {\em Phys. Rev. Lett.}, 59:1581--1584, October 1987.

\bibitem{HeldPhD}
M.~Held.
\newblock {\em Full-F gyro-fluid modelling of the tokamak edge and scrape-off
  layer}.
\newblock PhD thesis, 2016.

\bibitem{Held2018}
M.~Held, M.~Wiesenberger, R.~Kube, and A.~Kendl.
\newblock Non-oberbeck-boussinesq zonal flow generation.
\newblock {\em Nucl. Fusion}, 58(10):104001, 2018.

\bibitem{Held2016a}
M.~Held, M.~Wiesenberger, J.~Madsen, and A.~Kendl.
\newblock The influence of temperature dynamics and dynamic finite ion larmor
  radius effects on seeded high amplitude plasma blobs.
\newblock {\em Nucl. Fusion}, 56(12):126005, 2016.

\bibitem{Held2016}
M.~Held, M.~Wiesenberger, and A.~Stegmeir.
\newblock Three discontinuous galerkin schemes for the anisotropic heat
  conduction equation on non-aligned grids.
\newblock {\em Comput. Phys. Commun.}, 199:29--39, February 2016.

\bibitem{Higham02}
Nicholas~J. Higham.
\newblock {\em Accuracy and stability of numerical algorithms, second ed.}
\newblock Society for Industrial and Applied Mathematics (SIAM), Philadelphia,
  PA, 2002.

\bibitem{Iakymchuk15ExBLAS}
R.~Iakymchuk, S.~Collange, D.~Defour, and S.~Graillat.
\newblock {ExBLAS}: Reproducible and accurate {BLAS} library.
\newblock In {\em Proceedings of the Numerical Reproducibility at Exascale
  (NRE2015) workshop held as part of the Supercomputing Conference (SC15).
  Austin, TX, USA, November 15-20, 2015}, October 2015.

\bibitem{exblas}
R.~Iakymchuk, S.~Collange, D.~Defour, and S.~Graillat.
\newblock {ExBLAS (Exact BLAS) library}.
\newblock Available on the WWW, \url{https://exblas.lip6.fr/}, 2018.
\newblock Accessed 10-MAR-2018.

\bibitem{iakymchuk:hal-01419813}
R.~Iakymchuk, S.~Graillat, D.~Defour, and E.~S. Quintana-Ort{\'i}.
\newblock {Hierarchical Approach for Deriving a Reproducible LU factorization}.
\newblock Technical report, 2018.
\newblock Available on the WWW,
  \url{https://hal.archives-ouvertes.fr/hal-01419813}. Accessed 14-June-2018,
  HAL ID: hal-01419813.

\bibitem{blazeweb}
K.~Iglberger.
\newblock {Blaze}.
\newblock \url{https://bitbucket.org/blaze-lib/blaze}, 2018.

\bibitem{avx512}
Intel.
\newblock Intel\textsuperscript{\textregistered} architecture instruction set
  extensions programming reference.
\newblock Available on the WWW,
  https://software.intel.com/en-us/intel-architecture-instruction-set-extensions-programming-reference,
  2018.
\newblock Accessed 7-MAY-2018.

\bibitem{Kendl2014}
A.~Kendl.
\newblock Modelling of turbulent impurity transport in fusion edge plasmas
  using measured and calculated ionization cross sections.
\newblock {\em International Journal of Mass Spectrometry}, 365:106--113, May
  2014.

\bibitem{Kendl2017}
A.~Kendl, G.~Danler, M.~Wiesenberger, and M.~Held.
\newblock Interchange instability and transport in matter-antimatter plasmas.
\newblock {\em Phys. Rev. Lett.}, 118(23):235001, June 2017.

\bibitem{Knu97}
D.~E. Knuth.
\newblock {\em The Art of Computer Programming, Volume 2: Seminumerical
  Algorithms, third ed.}
\newblock Addison-Wesley, 1997.

\bibitem{Kube2016}
R.~Kube, O.~E. Garcia, and M.~Wiesenberger.
\newblock Amplitude and size scaling for interchange motions of plasma
  filaments.
\newblock {\em Phys. Plasmas}, 23(12):122302, 2016.

\bibitem{Kulisch11}
U.~Kulisch and V.~Snyder.
\newblock {The Exact Dot Product As Basic Tool for Long Interval Arithmetic}.
\newblock {\em Computing}, 91(3):307--313, March 2011.

\bibitem{Madsen2013}
J.~Madsen.
\newblock Full-f gyrofluid model.
\newblock {\em Phys. Plasmas}, 20(7):072301, July 2013.

\bibitem{Madsen2016}
J.~Madsen, V.~Naulin, A.~H. Nielsen, and J.~J. Rasmussen.
\newblock Collisional transport across the magnetic field in drift-fluid
  models.
\newblock {\em Phys. Plasmas}, 23(3):032306, March 2016.

\bibitem{meyers2005}
S.~Meyers.
\newblock {\em {Effective C++, Third Edition}}.
\newblock Addison-Wesley Professional, 2005.

\bibitem{meyers2014}
S.~Meyers.
\newblock {\em {Effective Modern C++}}.
\newblock O'Reilly Media, 2014.

\bibitem{HFPA10}
J.~M. Muller, N.~Brisebarre, F.~de~Dinechin, C.~P. Jeannerod, V.~Lef{\`e}vre,
  G.~Melquiond, N.~Revol, D.~Stehl{\'e}, and S.~Torres.
\newblock {\em Handbook of Floating-Point Arithmetic}.
\newblock Birkh\"auser, 2010.

\bibitem{numata07}
R.~Numata, R.~Ball, and R.~L. Dewar.
\newblock {Bifurcation in electrostatic resistive drift wave turbulence}.
\newblock {\em Phys. Plasmas}, 14:102312, 2007.

\bibitem{Ogita05accuratesum}
Takeshi Ogita, Siegfried~M. Rump, and Shin'ichi Oishi.
\newblock Accurate sum and dot product.
\newblock {\em SIAM J. Sci. Comput}, 26, 2005.

\bibitem{Rasmussen2016}
J.~J. Rasmussen, A.~H. Nielsen, J.~Madsen, V.~Naulin, and G.~S. Xu.
\newblock Numerical modeling of the transition from low to high confinement in
  magnetically confined plasma.
\newblock {\em Plasma Phys. Control. Fusion}, 58(1):014031, January 2016.

\bibitem{Scott2010}
B.~Scott.
\newblock {Derivation via free energy conservation constraints of gyrofluid
  equations with finite-gyroradius electromagnetic nonlinearities}.
\newblock {\em Phys. Plasmas}, 17(10):102306, October 2010.

\bibitem{Stegmeir2014}
A.~Stegmeir, D.~Coster, O.~Maj, and K.~Lackner.
\newblock Numerical methods for {3D} tokamak simulations using a flux-surface
  independent grid.
\newblock {\em Contrib. Plasma Phys.}, 54(4-6):549--554, June 2014.

\bibitem{Tamain2010}
P.~Tamain, P.~Ghendrih, E.~Tsitrone, V.~Grandgirard, X.~Garbet, Y.~Sarazin,
  E.~Serre, G.~Ciraolo, and G.~Chiavassa.
\newblock Tokam-{3D}: A {3D} fluid code for transport and turbulence in the
  edge plasma of tokamaks.
\newblock {\em J. Comput. Phys.}, 229(2):361--378, January 2010.

\bibitem{wakatani84}
M.~Wakatani and A.~Hasegawa.
\newblock {A collisional drift wave description of plasma edge turbulence}.
\newblock {\em Phys. Fluids}, 27:611, 1984.

\bibitem{WiesenbergerPhD}
M.~Wiesenberger.
\newblock {\em Gyrofluid computations of filament dynamics in tokamak
  scrape-off layers}.
\newblock PhD thesis, 2014.

\bibitem{performance}
M.~Wiesenberger.
\newblock {Feltor Performance Dataset}.
\newblock {\em Zenodo \url{http://doi.org/10.5281/zenodo.1290274}}, 2018.

\bibitem{Feltor-v5.0}
M.~Wiesenberger and M.~Held.
\newblock {Feltor v5.0}.
\newblock {\em Zenodo \url{http://doi.org/10.5281/zenodo.1290270}}, 2018.

\bibitem{wiesenberger2017}
M.~Wiesenberger, M.~Held, and L.~Einkemmer.
\newblock {Streamline integration as a method for two-dimensional elliptic grid
  generation}.
\newblock {\em J. Comput. Phys.}, 340:435--450, 2017.

\bibitem{Wiesenberger2018}
M.~Wiesenberger, M.~Held, L.~Einkemmer, and A.~Kendl.
\newblock Streamline integration as a method for structured grid generation in
  x-point geometry.
\newblock {\em J. Comput. Phys.}, (under review), 2018.

\bibitem{wiesenberger2017b}
M.~Wiesenberger, M.~Held, R.~Kube, and O~E Garcia.
\newblock Unified transport scaling laws for plasma blobs and depletions.
\newblock {\em Phys. Plasmas}, 24(6):064502, 2017.

\bibitem{wiesenberger2014}
M.~Wiesenberger, J.~Madsen, and A.~Kendl.
\newblock Radial convection of finite ion temperature, high amplitude plasma
  blobs.
\newblock {\em Phys. Plasmas}, 21:092301, 2014.

\bibitem{Wilkinson2016}
M.~D. Wilkinson, M.~Dumontier, I.~J. Aalbersberg, G.~Appleton, M.~Axton,
  A.~Baak, N.~Blomberg, J.~W. Boiten, L.~B.~D. Santos, P.~E. Bourne,
  J.~Bouwman, A.~J. Brookes, T.~Clark, M.~Crosas, I.~Dillo, O.~Dumon,
  S.~Edmunds, C.~T. Evelo, R.~Finkers, A.~Gonzalez-Beltran, A.~J.~G. Gray,
  P.~Groth, C.~Goble, J.~S. Grethe, J.~Heringa, P.~A.~C. 't~Hoen, R.~Hooft,
  T.~Kuhn, R.~Kok, J.~Kok, S.~J. Lusher, M.~E. Martone, A.~Mons, A.~L. Packer,
  B.~Persson, P.~Rocca-Serra, M.~Roos, R.~van Schaik, S.~A. Sansone,
  E.~Schultes, T.~Sengstag, T.~Slater, G.~Strawn, M.~A. Swertz, M.~Thompson,
  J.~van~der Lei, E.~van Mulligen, J.~Velterop, A.~Waagmeester, P.~Wittenburg,
  K.~Wolstencroft, J.~Zhao, and B.~Mons.
\newblock Comment: The fair guiding principles for scientific data management
  and stewardship.
\newblock {\em Scientific Data}, 3:UNSP 160018, March 2016.

\bibitem{Zeiler1997}
A.~Zeiler, J.~F. Drake, and B.~Rogers.
\newblock {Nonlinear reduced Braginskii equations with ion thermal dynamics in
  toroidal plasma}.
\newblock {\em Phys. Plasmas}, 4(6):2134--2138, 1997.

\end{thebibliography}
